\documentclass[preprint,amsmath,amssymb,superscriptaddress,longbibliography,aps,floatfix,footinbib]{revtex4-2}

\usepackage{graphicx}
\usepackage{tabularx}
\usepackage[usenames,dvipsnames]{color}
\usepackage{soul}
\usepackage{bm}
\usepackage{amsmath}
\usepackage{gensymb}

\usepackage{times}
\usepackage{amsfonts}
\usepackage{mathrsfs}
\usepackage{dcolumn}
\usepackage{bm}
\usepackage{color}

\usepackage[colorlinks,bookmarks=false,citecolor=blue,linkcolor=red,urlcolor=blue]{hyperref}
\usepackage{multirow}
\usepackage{physics}
\usepackage{siunitx}
\DeclareSIUnit\angstrom{\text {Å}}

\hypersetup{%
    ,urlcolor=black
    ,citecolor=black
    ,linkcolor=black
    }

\newcommand\quotemarks[1]{‘‘#1’’}

\begin{document}

\title{Termination-dependent surface Fermiology in the intercalated altermagnet CoNb$_4$Se$_8$}
\author{Pietro Maria Forcella$^\$$}
\affiliation{Université Paris-Saclay, CNRS, Centre de Nanosciences et de Nanotechnologies 91120, Palaiseau, Paris, France}

\author{Alessandro De Vita$^\$$}
\affiliation{Fritz Haber Institut der Max Planck Gesellschaft, Faradayweg 4--6, 14195 Berlin, Germany\looseness=-1}

\author{Matthew D. Watson}
\affiliation{Diamond Light Source Ltd, Harwell Science and Innovation Campus, Didcot, OX110DE, United Kingdom}

\author{Sammar Tayyab}
\affiliation{Dipartimento di Fisica, Sapienza Università di Roma, P.le Aldo Moro 2, 00185 Rome, Italy}

\author{Chiara Bigi}
\affiliation{Synchrotron SOLEIL, L’Orme des Merisiers, D\'epartementale 128, F-91190 Saint-Aubin, France}

\author{Iulia Cojocariu}
\affiliation{Elettra - Sincrotrone Trieste S.C.p.A., Strada Statale 14, km 163.5, 34149 Trieste, Italy}
\affiliation{Dipartimento di Fisica, Università degli Studi di Trieste, 34127, Trieste, Italy}

\author{Marcin Szpytma}
\affiliation{Elettra - Sincrotrone Trieste S.C.p.A., Strada Statale 14, km 163.5, 34149 Trieste, Italy}
\affiliation{Faculty of Physics and Applied Computer Science, AGH University of Krakow, Władysława Reymonta 19, 30-059 Krakow, Poland}

\author{Matteo Jugovac}
\affiliation{CNR-Istituto di Struttura della Materia (CNR-ISM),
Strada Statale 14, km 163.5, 34149 Trieste, Italy}

\author{Andrea Locatelli}
\affiliation{Elettra - Sincrotrone Trieste S.C.p.A., Strada Statale 14, km 163.5, 34149 Trieste, Italy}

\author{Tevfik Onur Menteş}
\affiliation{Elettra - Sincrotrone Trieste S.C.p.A., Strada Statale 14, km 163.5, 34149 Trieste, Italy}

\author{My Bang Petersen}
\affiliation{Department of Physics and Astronomy, Interdisciplinary Nanoscience Center, Aarhus University, 8000 Aarhus C, Denmark}

\author{Paolo Moras}
\affiliation{CNR-Istituto di Struttura della Materia (CNR-ISM),
Strada Statale 14, km 163.5, 34149 Trieste, Italy}

\author{Polina M. Sheverdyaeva}
\affiliation{CNR-Istituto di Struttura della Materia (CNR-ISM),
Strada Statale 14, km 163.5, 34149 Trieste, Italy}

\author{Christopher Candelora}
\affiliation{Department of Physics, Boston College, Chestnut Hill, MA 02467, USA}

\author{Ilija Zeljkovic}
\affiliation{Department of Physics, Boston College, Chestnut Hill, MA 02467, USA}

\author{Federico Motti}
\affiliation{CNR-Istituto Officina dei Materiali (IOM), Unità di Trieste, Strada Statale 14, km 163.5, 34149 Basovizza (TS), Italy\looseness=-1}%

\author{Jill A. Miwa}
\affiliation{Department of Physics and Astronomy, Interdisciplinary Nanoscience Center, Aarhus University, 8000 Aarhus C, Denmark}

\author{Vincent Polewczyk}
\affiliation{Universit\'e Paris-Saclay, UVSQ, CNRS, GEMaC, 45 Av. des États Unis, 78000 Versailles, France}

\author{Matthieu Jamet}
\affiliation{Univ. Grenoble Alpes, CEA, CNRS, IRIG-SPINTEC, 17 Av. des Martyrs, 38054 Grenoble, France}



\author{Matteo Calandra}\email{m.calandrabuonaura@unitn.it}
\affiliation{Department of Physics, University of Trento, Via Sommarive 14, Povo 38123, Italy}

\author{Younghun Hwang}\email{younghh@uc.ac.kr}
\affiliation{Center for Artificial Low-Dimensional Electronic Systems, Institute for Basic Science (IBS), Pohang 37673, Republic of Korea}
\affiliation{Electricity \& Electronics and Semiconductor Engineering, Ulsan College, Ulsan 44610, Republic of Korea}

\author{Davide Romanin}\email{davide.romanin@universite-paris-saclay.fr}
\affiliation{Université Paris-Saclay, CNRS, Centre de Nanosciences et de Nanotechnologies 91120, Palaiseau, Paris, France}

\author{Federico Mazzola}\email{federico.mazzola@unipd.it}
\affiliation{Department of Physics and Astronomy ‘Galileo Galilei’, University of Padova, Via Francesco Marzolo, 8, 35121 Padova, Italy}
\affiliation{CNR-Istituto di Struttura della Materia (CNR-ISM),
Strada Statale 14, km 163.5, 34149 Trieste, Italy}

\begin{abstract}
\bf{Altermagnets have recently gained attention for hosting symmetry-enforced spin-split bands despite zero net magnetization. Angle-resolved photoelectron spectroscopy (ARPES) offers direct access to their electronic structure. However, the intrinsic surface sensitivity of this technique complicates the identification of the observed features: Surface states can be misinterpreted as the spin-split counterpart of bulk bands. This issue becomes particularly relevant for layered intercalated systems, where micrometer-sized domains with distinct terminations often coexist on a cleaved surface. We discuss this scenario for CoNb$_4$Se$_8$, one of the first established layered altermagnets. Supported by density functional theory calculations, we employ micro-ARPES to probe individual domains and reveal two contrasting surface terminations. We identify Co- and Se-terminated domains with characteristic surface states arising from bulk–surface connectivity of the material. Our work shows that termination-dependent surface contributions are essential for the reliable spectroscopic identification of altermagnetism and for assessing its presence across the family of layered materials.}
\end{abstract}

\maketitle
\noindent
$^\$$P.M.F. and A.D.V. contributed equally to this work.


Altermagnetism has recently emerged as a fundamental class of magnetism in which specific crystalline symmetries enforce robust spin-split bulk electronic bands despite zero net magnetization \cite{Smejkal2022_1, Smejkal2022_2, Mazin2022,Bai2024}. This effect was first proposed in theoretical works predicting non-relativistic momentum-dependent spin polarisation in antiferromagnets \cite{Smejkal2022_1,Turek2022,Yuan2023}, and later observed experimentally in material candidates such as MnTe, RuO$_2$, and CrSb \cite{Mazin2023, Feng2022,Krempasky2024, Fedchenko2024}. Angle-Resolved Photoelectron Spectroscopy (ARPES) is a powerful tool for uncovering such unconventional spin splittings in real materials \cite{Liu2026}. However, the interpretation of ARPES measurements is often complicated by surface-related contributions, as exemplified by the case of RuO$_2$ \cite{Fedchenko2024, Lin2024, Liu2024}. 

Co-intercalated NbSe$_2$ compounds have attracted interest for exhibiting symmetry conditions compatible with an altermagnetic band structure. CoNb$_4$Se$_8$, composed of Co atoms intercalated between NbSe$_2$ layers, hosts a collinear antiferromagnetic ground state and has been proposed as one of the first realisations of altermagnetism in layered materials \cite{Regmi2024, Devita2026, Candelora_2026}. Electronic-structure calculations predict sizeable spin splitting and a highly anisotropic spin texture, and spectroscopic measurements report bulk-like features showing non-zero spin polarisation at multiple \textit{k}-points, consistent with an altermagnetic electronic structure \cite{Devita2026}. While previous studies focused on bulk properties, surface effects are expected to play a crucial role in altermagnets. The symmetry-enforced, momentum-dependent spin splitting of the bulk can be lost at the surface \cite{Sattigeri2023}; specifically, a single-layer projection of the electronic structure of CoNb$_4$Se$_8$ along the \textit{c} axis ([001] direction) implies $k_z=0$ and thus a nodal plane, enforcing spin-degenerate states in the whole Brillouin zone. In addition, surface states can obscure bulk bands \cite{Sattigeri2023, Mazzola_2023_co, DiSante_2023}. Despite the possibility of themselves mediating exotic phases \cite{Mazzola_2024,DeVita2026b}, in this context surface states challenge the clear identification of bulk altermagnetic features in the material \cite{Edwards2023}. 

\begin{figure}
\includegraphics[width=0.8\columnwidth]{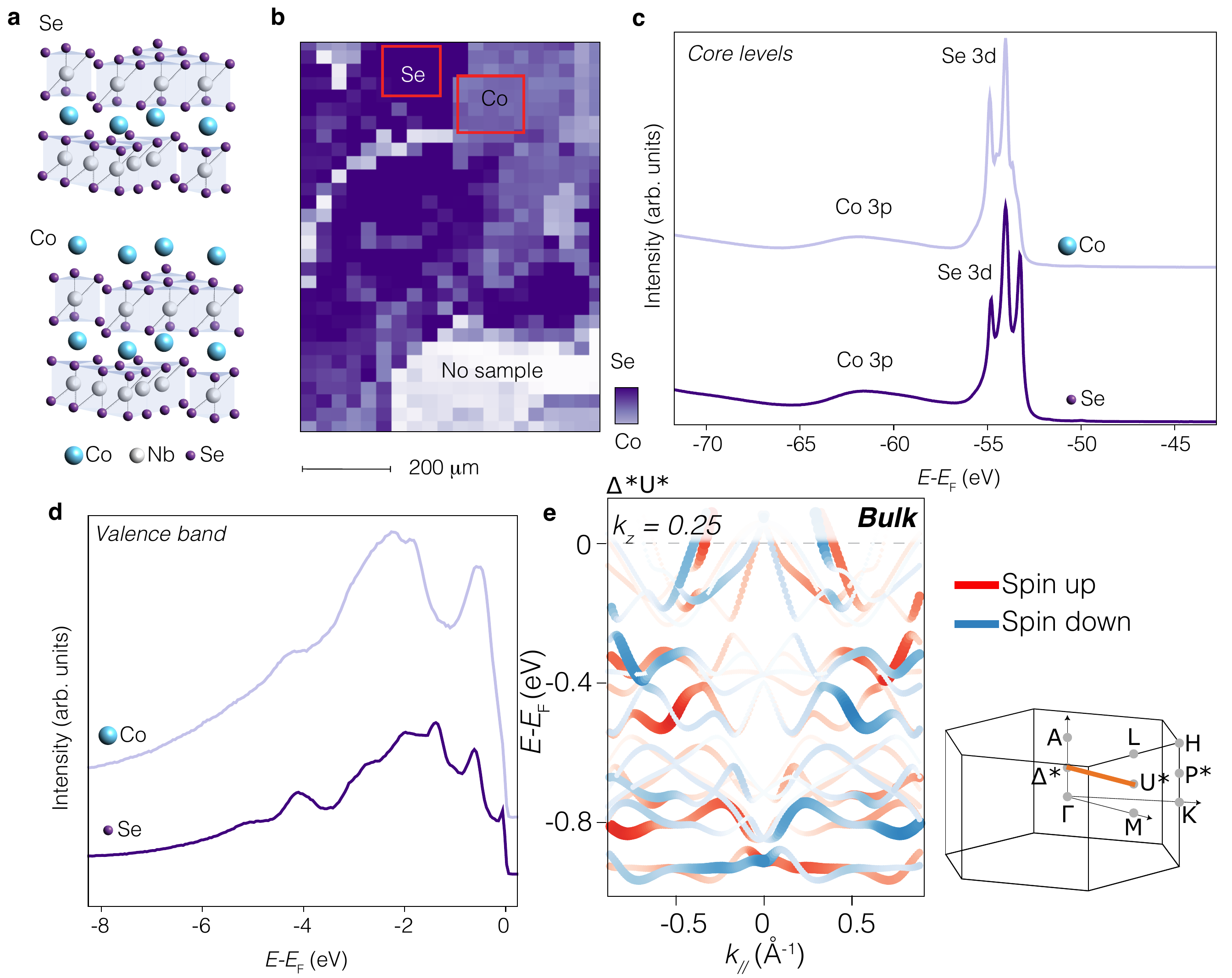}
  \caption{\textbf{a.} Schematics of the crystal structure and the possible surface terminations obtained after cleaving the sample. \textbf{b.} Micro-ARPES spatial map obtained by raster-scanning the beam and integrating, at each pixel, the valence-band intensity over the termination-sensitive window $E\in[43.0,43.13]~\mathrm{eV}$ and $k\in[-1.85,-0.85]~\mathrm{\AA}^{-1}$. This window was chosen from the single-domain spectra and slab-DFT fingerprints and is used as a spatial guide only; the final termination assignment is made spot-by-spot from the agreement between the core-level lineshape and the near-$E_F$ band fingerprint, without applying an automated intensity-ratio threshold. Here for simplicity we employ deep-purple for Se, light purple for Co, and white for absence of signal. \textbf{c.} The distinct surface character is supported also by micro-resolved core-level X-ray photoelectron spectroscopy on the same spots acquired at \SI{120}{\eV} photon energy, showing features reconcilable with Co and Se surface spectra and in accordance with previous works. \textbf{d.} Valence band photoemission spectroscopy signal for the two terminations, acquired at \SI{120}{\eV} photon energy. As detailed in the main text, the differences are reflected by an overall surface doping due to the presence of a different termination and the presence of surface states. \textbf{e.} For completeness, we show consistently with Ref.\cite{Devita2026}, the calculated bulk electronic structure along the direction indicated in the Brillouin zone, featuring a spin-pattern consistent with altermagnetism.}
  \label{fig:1}
\end{figure}

Moreover, some experimental challenges require serious attention. For example, cleaving the intercalated CoNb$_4$Se$_8$ can expose inequivalent atomic planes, \textit{i.e.}, chalcogen- versus intercalant-terminated surfaces, each characterised by a distinct surface potential and hosting its own set of surface bands. When these termination-dependent states disperse near the bulk altermagnetic bands, they can modify the apparent bulk–surface connectivity and be misinterpreted as additional spin-split branches, thereby distorting the interpretation of the inferred electronic structure. The importance of this effect is exemplified by the magnetic Weyl kagome metal Co$_3$Sn$_2$S$_2$, where micro-ARPES and first-principles calculations have shown that different surface terminations host distinct topological surface states and even alter the connectivity of Weyl points while leaving the underlying bulk nodes intact \cite{Mazzola_2023_co}. Standard ARPES is also intrinsically limited in resolving termination-dependent electronic structures in layered materials: typical termination domains extend over only a few tens of microns \cite{Mazzola_2018, Mazzola_2023_co}, thus the photon footprint inevitably averages over multiple surface regions, and spectral broadening masks the distinct dispersions associated with individual terminations. Such averaging can wash out termination-specific states or merge them into apparent bulk features, making it challenging to disentangle genuine altermagnetic band splittings from surface-derived replicas. These limitations are particularly severe for CoNb$_4$Se$_8$, whose structure naturally permits both chalcogen and intercalant terminated surfaces (see Fig.~\ref{fig:1}a) \cite{Devita2026, Candelora_2026}. As a result, conventional spectroscopies cannot determine whether observed spectral features originate from the intrinsic bulk altermagnetic band structure or instead from surface-localised states associated with one or more terminations. Despite these precedents, the termination-dependent surface electronic structure of layered altermagnets such as CoNb$_4$Se$_8$ has not yet been characterized.

To overcome these challenges, we employed micro-resolved experimental techniques for the characterization of the surface electronic properties of CoNb$_4$Se$_8$. Micro-ARPES allowed us to probe selectively two single-surface terminations, \textit{i.e.}, Co and Se. By reducing the probing area well below the characteristic domain size, our measurements granted direct access to the termination-dependent electronic structure of CoNb$_4$Se$_8$ without artefacts arising from spatial averaging. The excellent correspondence of experimental data with first-principle Density Functional Theory (DFT) calculations fully proves the attribution to the Co- and Se-terminated areas. Our results provide a comprehensive overview of the surface contribution to the Fermiology in CoNb$_4$Se$_8$, while also establishing a methodological precedent for robust distinction of bulk altermagnetic features from surface-derived bands.  

Single crystals of CoNb$_4$Se$_8$ were grown by chemical vapor transport following procedures described in our previous works on Co-intercalated NbSe$_2$ \cite{Devita2026,Candelora_2026}, where we demonstrated that controlled Co intercalation between NbSe$_2$ layers leads to a doubling of the crystallographic unit cell, producing a robust $2\times2$ modulation visible across a broad temperature range. Some of the photoemission-resolved datasets analyzed here were acquired as part of the experimental campaign reported in Ref. \cite{Devita2026}. In some cases, measurements were performed on the same samples, on different cleaves. Full synthesis details and characterisation procedures for the present crystals are reported in the Methods section of Ref. \cite{Devita2026}, and in the Supporting Information. Micro-ARPES measurements were performed at the I05-Nano branch of Diamond Light Source (DLS), providing spectra with a lateral spatial resolution better than \SI{4}{\um}. The full description of the DFT calculations methods \cite{QEcode,QE-2017,QE-2020,PBE,Methfessel1989,PhysRevB.85.085201,Pacil2021} can be found in the Supporting Information.

We stress that the termination assignment used here does not contradict the real-space STM periodicities reported previously for CoNb$_4$Se$_8$ \cite{Devita2026,Candelora_2026}. The $2a_0\times2a_0$ modulation is a structural consequence of the ordered Co intercalation, which doubles the in-plane unit cell of the parent $2H$-NbSe$_2$, and is therefore common to both cleavage terminations rather than being a unique fingerprint of only one surface. Its STM contrast and Fourier-transform amplitude can differ strongly between terminations because STM probes either a projected modulation of the outer Se lattice, with Fourier components at half the Se Bragg wavevector, or the more direct residual ordered Co ad-lattice left on the surface after cleaving. Within this picture, additional Fourier features reported previously reflect termination-dependent tunnelling matrix elements, surface sensitivity, and local disorder, rather than a structural inconsistency with the present assignment.

We first establish the existence of two distinct surface terminations in CoNb$_4$Se$_8$. This is directly visualised using micro-ARPES, as shown in Fig.~\ref{fig:1}b, where photoemission intensity maps reflect the chemical heterogeneity present on the surface. At each spatial position, Se 3\textit{d} and Co 3\textit{p} core-levels are monitored, enabling the identification of distinct surface features reconcilable with surface-localized Co and Se, as illustrated in Fig.~\ref{fig:1}c for representative regions. From core level analysis (See Supporting Information Fig.~\ref{fig:S1} for fits and quantitative details), the investigated areas show a variation in the relative Co vs. Se signal, consistent with distinct terminations. Se-terminated regions exhibit additional peaks due to surface core levels shift, while in the Co-terminated regions such replicas are significantly suppressed in intensity. The two terminations also exhibit markedly different valence-band spectra: angle-integrated measurements (Fig.~\ref{fig:1}d) show distinct lineshapes consistent with different surface potentials, while both terminations retain metallic character with bands crossing the Fermi level. The density of states near $E_F$ has mixed atomic character, with termination-dependent Nb, Se, and Co contributions; the Se and Co components are larger for the corresponding termination, as shown by the atom-projected PDOS in the Supporting Information (Fig.~\ref{fig:pdos_terminations}).

\begin{figure}
\includegraphics[width=0.65\columnwidth]{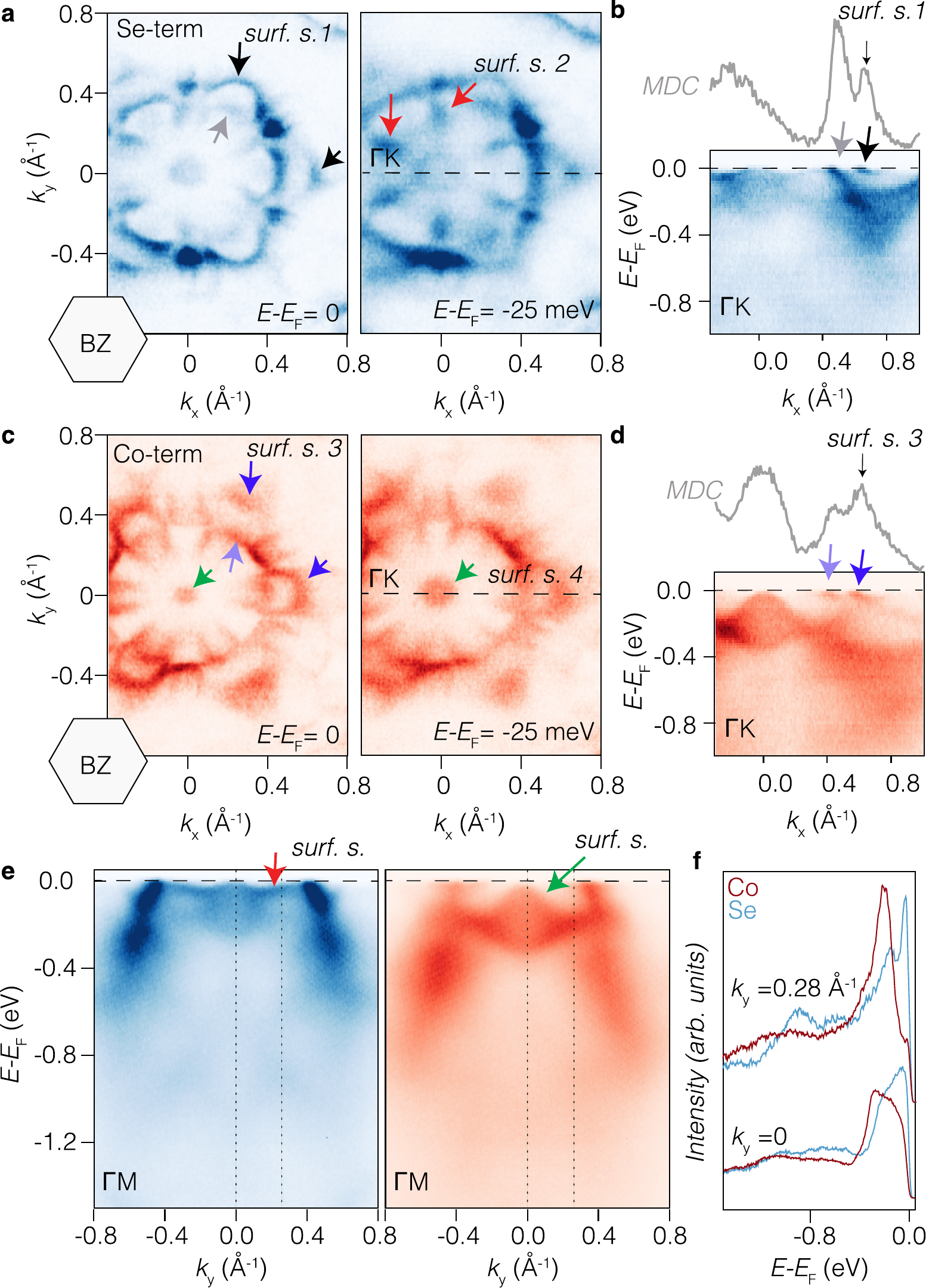}
  \caption{Fermi surface (\textit{left}) and constant energy contour at $E-E_F=\SI{-25}{\milli\eV}$ (\textit{right}) for the \textbf{a.} Se and \textbf{c.} Co terminations. The intensity shows a threefold symmetric pattern with elongated petals. The arrows indicate termination-dependent surface-derived features or surface resonances; same colors track the same features across the spectra. Such states are prominent especially along the $\Gamma$–K directions, shown in \textbf{b.} for Co and \textbf{d.} for Se. Along with the ARPES spectra, momentum distribution curves (MDCs) at the Fermi level are reported. \textbf{e.} Spectra along $\Gamma$–M for both terminations are compared. While there are near-Fermi level features visible in the Se-termination, these are absent in the Co one. An overall $\sim\SI{40}{\milli\eV}$ shift of the Fermi energy is also observed consistent with electron doping provided by the presence of Co at the top surface. However, even by shifting the Fermi level down by \SI{60}{\milli\eV} for the Co-termination spectra, the resulting energy distribution curves extracted at different momenta (dotted vertical lines in \textbf{e.}), shown in \textbf{f.}, still look significantly different (see also the comparison with constant energy maps in Supporting Information, Fig.~\ref{fig:S2}).}
  \label{fig:2}
\end{figure}

However, a quantification by means of (angle-integrated) XPS carries a large uncertainty, associated both with the inelastic mean free path and with the possibility of additional Se features tailing into the Co region, the relevant peaks being very close to one another; XPS therefore is able to provide only circumstantial evidence and is intended to support, rather than establish, the two-termination picture. For this reason, to elucidate the surface-state Fermiology and hereby provide decisive evidence for the attribution of surface terminations, we acquired both Fermi surface maps and high-resolution energy–momentum maps using micro-ARPES, focusing exclusively on well-ordered regions exposing the features sketched in Fig.\ref{fig:1}. We therefore label these measurements accordingly as "Se-term" and "Co-term". The Fermi surfaces (shown in Fig.~\ref{fig:2}a in blue and Fig.~\ref{fig:2}c in red, respectively) both exhibit flower-like patterns, yet they differ markedly already at first inspection; for example, they feature different shapes in the Fermi surface. Notably, these datasets were acquired at the same photon energy, polarization, and experimental conditions, ensuring a direct and reliable comparison. For both situations, the Fermi surfaces do not share electronic structures similar to the undoped parent compound NbSe$_2$ \cite{Bawden_2016, Bigi_2021}, but the Co intercalation drastically modifies the Fermiology.

The constant energy surface pattern becomes even more evident in constant-energy maps extracted below the Fermi level, as illustrated by the data at \SI{25}{\milli\eV} binding energy (BE) in Fig.~\ref{fig:2}a-b. For constant energy maps at additional binding energies, see also the data presented in Supporting Information Fig.~\ref{fig:S2}. The termination-dependent differences are further clarified by analysing momentum-resolved photoemission spectra along the high-symmetry directions $\bar{\Gamma}-\bar{K}$ and $\bar{\Gamma}-\bar{M}$, as shown in Fig.~\ref{fig:2}b–e. For both terminations, sharp dispersive features are observed along $\bar{\Gamma}-\bar{K}$ and cross the Fermi level. These states are clearly separated from the broader valence-band spectral weight and are labeled \quotemarks{surf.~s. 1-3} in the figure, additionally they accounts for the arc-like intensity observed in the corresponding Fermi-surface maps (Fig.~\ref{fig:2}a,c).

Additional termination-dependent features become evident in constant-energy maps acquired at finite binding energies. At $E-E_F=\SI{-25}{\milli\eV}$, the Se-terminated surface exhibits elongated intensity segments arranged with an overall sixfold symmetry, whereas no analogous pattern is observed for the Co-terminated surface. Consistently, the dispersions measured at $E-E_F=\SI{-25}{\milli\eV}$ on the Se termination show minimally dispersing bands along $\bar{\Gamma}-\bar{M}$ in a \quotemarks{cartwheel spoke} fashion (\quotemarks{surf. s. 2}) around a central \quotemarks{bore}-like pocket at $\bar{\Gamma}$. These states are absent for the Co-terminated surface (Fig.~\ref{fig:2}e). The contrast between the two terminations is further emphasised by energy distribution curves (EDCs) extracted at selected momenta (Fig.~\ref{fig:2}f), which reveal clear differences in both peak positions and spectral weight.

We point out that the differences between the band texture of the two terminations cannot be explained solely by local doping variations across the surface. This is demonstrated by analysing a series of constant-energy maps (Supporting Information, Fig.~\ref{fig:S2}). The central $\bar{\Gamma}$-centred pocket is not used here to assign the termination-dependent states, but only as a common registration feature to test the rigid-shift hypothesis; its precise bulk, surface, or surface-resonance character is not essential to our conclusions. Matching its area requires a finite energy shift between the two terminations, confirming a local chemical-potential offset; however, once this shift is applied, the remaining petal- and spindle-like features still do not overlap in shape, connectivity, or spectral weight. The two Fermiologies therefore cannot be explained by a rigid shift alone, and instead reflect distinct termination-dependent surface or surface-resonance contributions.

\begin{figure}
\includegraphics[width=1\columnwidth]{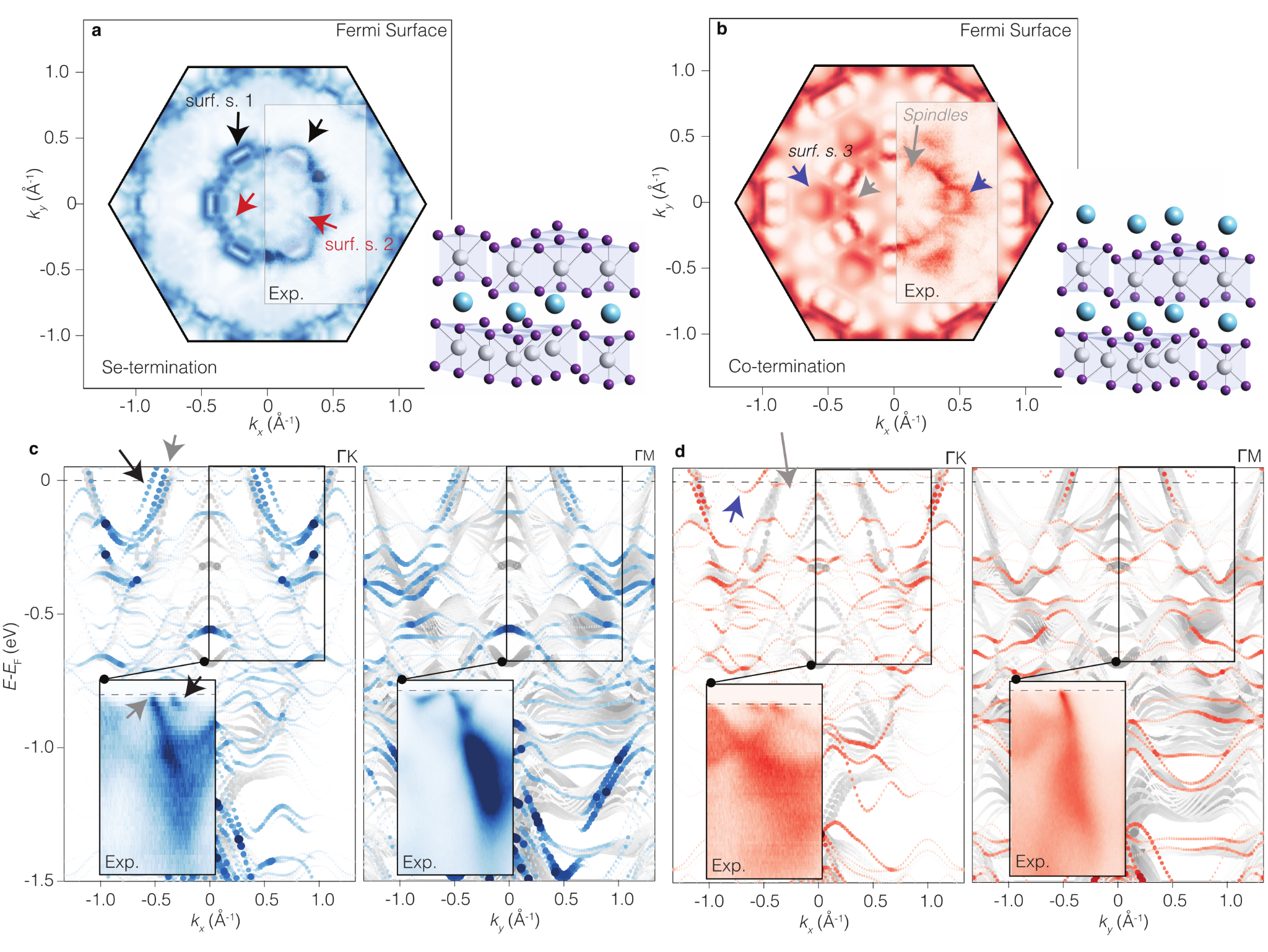}
  \caption{ Calculated Fermi surfaces for the (a) Se- and (b) Co-terminated slabs, with the corresponding experimental spectra overlaid in the inset. Both terminations exhibit petal-like pockets (black arrows) and spindle-shaped termination-dependent features (red/green arrows) that form a manifold of surface-derived states and surface resonances, predominantly distributed along the $\Gamma$–K directions. No additional bulk-derived bands are found to cross the Fermi level along $\Gamma$–K in the same \textit{k}-space region, consistent with the symmetry-constrained band structure. (c) Calculated energy–momentum dispersions for the Se-terminated surface along the $\Gamma$–K and $\Gamma$–M high-symmetry directions, and (d) the corresponding dispersions for the Co termination. While the petal and spindle features are clearly resolved along $\Gamma$–K for both terminations, termination-dependent differences emerge at finite binding energies. Features spectrally separated from the projected bulk continuum are classified as surface-localised states, whereas termination-dependent features overlapping the continuum are described as surface resonances; both categories are consistent with the experimental observations reported in Fig.~\ref{fig:2}.}
  \label{fig:3}
\end{figure}

The experimental observations described above find direct correspondence in density functional theory (DFT) calculations (Fig.~\ref{fig:3}), which explicitly account for the two possible surface terminations. By modeling slab geometries with Se- and Co-terminated surfaces, we compute both the Fermi surfaces and the momentum-resolved photoemission spectra. Details of the calculations are reported in the Supporting Information, along with bulk-projected and slab spectra plotted separately for both terminations and along both $\Gamma$-M and $\Gamma$-K (Fig.~\ref{fig:S3}). In the assessment of the DFT-calculated spectra, we classify a feature as an isolated termination-derived surface state only when two criteria are satisfied: (i) it lies in an energy--momentum region where no bulk band is present in the bulk projection, \textit{i.e.}, outside the grey bulk continuum, and (ii) its dispersion and spectral weight depend on the termination, appearing for one termination while being absent, or different, for the other. When a termination-dependent feature overlaps the projected bulk continuum, we do not classify it as an isolated surface state but describe it as a surface resonance. This distinction is supported by the atom- and layer-projected DOS shown in the Supporting Information (Figs.~\ref{fig:pdos_terminations} and \ref{fig:surface_vs_bulk_pdos}). The calculations accurately reproduce the key experimental features, including the petal-like Fermi-surface contours (black arrows), the spindle-shaped states (red arrows for Se and grey arrows for Co), and the pronounced termination-dependent differences below the Fermi level (Fig.~\ref{fig:3}a-b). In particular, surface-derived bands appear selectively for one termination while remaining absent for the other, consistent with the termination-specific states identified by micro-ARPES. The same petal- and spindle-like features are also evident in the calculated energy–momentum dispersions of Fig.~\ref{fig:3}c,d along the $\Gamma$-K direction, where well-defined surface-derived bands approach and intersect the Fermi level, in agreement with the arc-like intensity observed experimentally. By contrast, along $\Gamma$-M no states with clear surface character cross $E_F$ at similar \textit{k}-values, indicating that this direction does not contribute prominently to the petal surface-state manifold at the Fermi surface. Nevertheless, a closer inspection of the dispersions at lower \textit{k}-values reveals a rich termination-dependent structure also along $\Gamma$-M. In particular, for the Se-terminated surface two weakly dispersive bands appear just below $E_F$, forming locally flat features encompassed by the bulk states, whereas the Co-terminated surface exhibits a distinct diamond-shaped state shifted to lower binding energy, as highlighted in Fig.~\ref{fig:3}c-d. In particular we note that, for both terminations, along $\Gamma$-M surface-derived contributions disperse closely alongside the bulk bands. This proximity identifies these overlapping features as surface resonances and shows how surface-derived spectral weight can obscure or mimic bulk altermagnetic features when contributions from different terminations are spatially averaged. On the other hand, micro-ARPES disentangles the tiny termination-induced spectral weight shifts, and can therefore univocally select one termination where bulk altermagnetic features are not mixed with multiple surface Se- and Co-induced contributions. 

Note that the experimental electronic structure appears broader and many features visible in the DFT are smeared out. This is likely due to a combination of photoemission matrix elements, a possible contribution from the bulk electronic structure (also shown in Fig.~\ref{fig:3} in the DFT with the gray shadowing), and presence of local defects and vacancies. Additionally, the broadening is further enhanced by $k_z$ effects, which will necessarily give spectral contribution from a finite portion of the Brillouin zone, as discussed in the Supporting Information. In any case, the strong correspondence between data and calculations provides compelling evidence for the definitive attribution of Co and Se terminations.

\begin{figure}
\includegraphics[width=1\columnwidth]{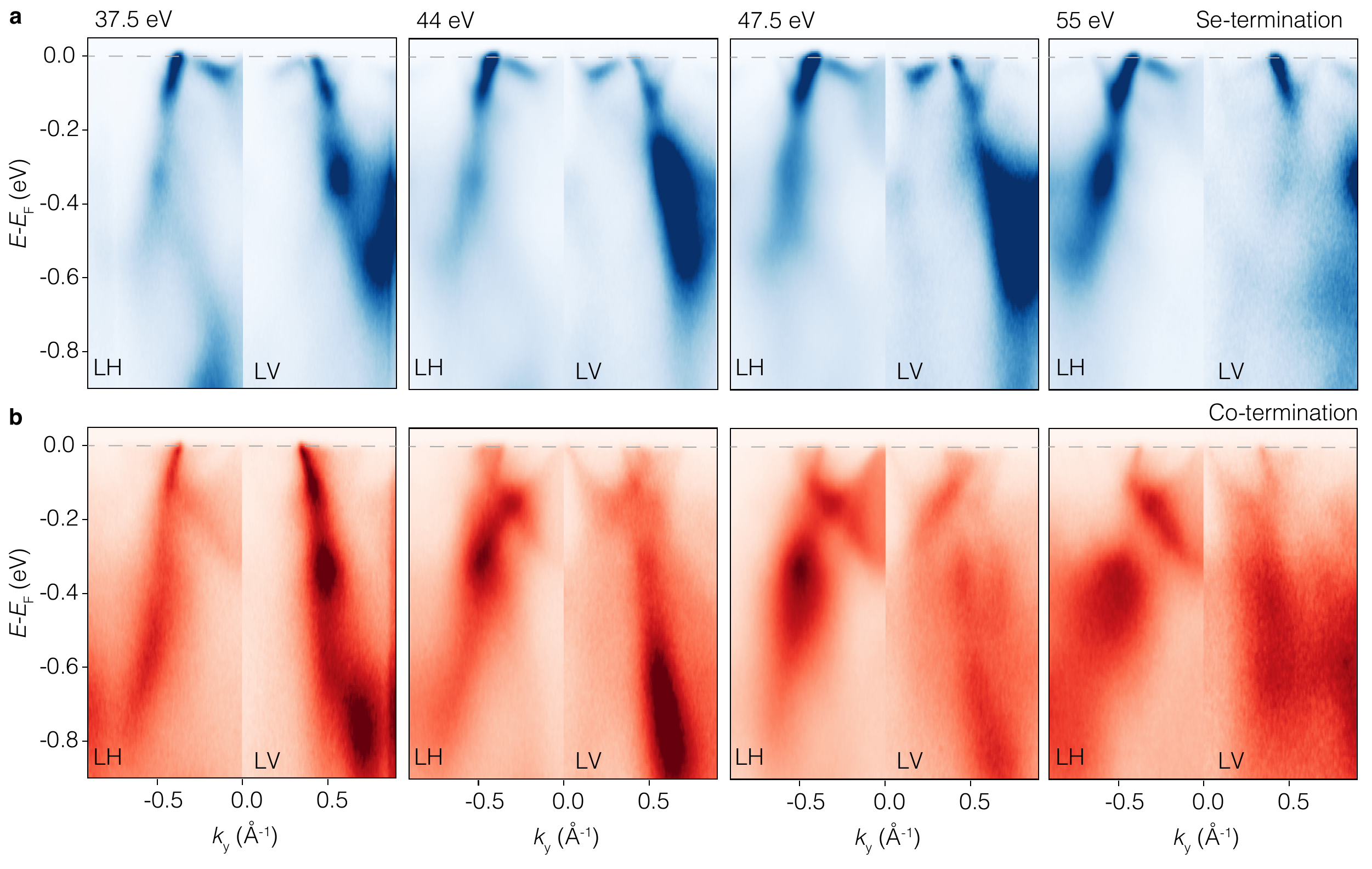}
  \caption{Energy-momentum photoemission spectra along $\Gamma$–M as a function of photon energy, for the Se- and Co-terminated surfaces. Measurements are performed with linearly polarized light along the vertical (\textit{s}-polarization) and horizontal (\textit{p}-polarization) directions, to address bands with different symmetries.}
  \label{fig:4}
\end{figure}

To experimentally corroborate the surface state character of the observed features, we measured energy–momentum photoemission spectra along the $\Gamma$-M direction for both surface terminations under a range of photon energies and light polarizations. Representative datasets are shown in Fig.~\ref{fig:4}, where the Se-terminated surface exhibits weakly dispersive states just below $E_F$, and the Co-terminated surface shows a deeper, diamond-shaped feature at lower binding energies, fully consistent with the DFT results shown in Fig.~\ref{fig:3}c-d. No additional bands emerge, nor do the observed features display significant changes in dispersion or intensity upon varying either photon energy or polarization. This weak dependence suggests a mixed in-plane and out-of-plane orbital character of the states and, importantly, the absence of any appreciable $k_z$ dispersion, supporting their predominantly two-dimensional, surface-derived or surface-resonance character.

In conclusion, we demonstrated that the cleaved surface of CoNb$_4$Se$_8$ intrinsically hosts two inequivalent terminations, each giving rise to a distinct manifold of surface-derived electronic states and surface resonances. By combining micro-ARPES with micron-scale spatial selectivity and density functional theory calculations, we directly disentangle the electronic structures associated with the Se- and Co-terminated areas and show that their Fermi surfaces, dispersions, and low-energy states differ qualitatively rather than by a simple rigid chemical-potential shift. Termination-specific bands appear depending on the exposed atomic layer, producing petal- and spindle-like features along $\Gamma$–K and additional flat or diamond-shaped states along $\Gamma$–M that are confirmed by both theory and experiment. The weak photon-energy and polarization dependence establishes the predominantly two-dimensional surface-derived character of these states.

These results highlight that in layered altermagnets, surface contributions are not a secondary effect but a crucial element of the measured photoemission spectra. We note that the presence of surface states does not alter the conclusions regarding the emergence of altermagnetism in this compound \cite{Regmi2024,Devita2026}. However, without spatially resolved measurements, the coexistence of multiple terminations would lead to spectral averaging and could mimic or obscure the symmetry-driven spin splittings that characterise altermagnetism. Our work therefore provides both a microscopic description of the surface electronic structure of CoNb$_4$Se$_8$ and a general methodological framework for reliably separating bulk altermagnetic bands from termination-dependent surface-derived states and surface resonances. This approach will be essential for the robust spectroscopic identification of altermagnetism across the growing family of layered compounds and for accurately reconstructing their intrinsic Fermiology.

\noindent{\bf{\large Acknowledgments}}\\
F.M. acknowledges MUR funding within the FIS2 (n. 1236, 1-8-2023) Project no. FIS-2023-03652 (CUP D53C25001950001). A.D.V. acknowledges the Deutsche Forschungsgemeinschaft (DFG) within Transregio TRR 227 Ultrafast Spin Dynamics project B07, the Max Planck Society, and BERLIN QUANTUM, an initiative endowed by the Innovation Promotion Fund of the city of Berlin. J.A.M and M.B.P. gratefully acknowledge support from the Danish Agency for Science and Higher Education (DanScatt Grants No. 7129-00018B and 7129-00026B). Y.H. was supported by the Institute for Basic Science (IBS) through the Center for Artificial Low-Dimensional Electronic Systems (IBS-R014-D1) and by the National Research Foundation of Korea (NRF) grant funded by the Korean government (MSIT) (RS-2025-24742993). The authors gratefully thank Prof. Maria Grazia Betti, Prof. Carlo Mariani, and Dr. Giovanni Vinai for preliminary measurements and fruitful discussions.

\noindent{\large \bf{Contributions}}\\
F.M., Y.H., and D.R. conceived the project. F.M., A.D.V., M.D.W, S.T., C.B., I.C., M.S., A.L., T.O.M., M.J., M.B.P., J.A.M, P.M., P.M.S., I.J., C.C., F.Mo., G.V., V.P., M.Ja. performed the experimental part. M.C., P.F., and D.R. performed the DFT calculations and Y.H. performed synthesis and characterization of the crystals. All authors contributed writing the manuscript.

\noindent{\bf{\large Ethics declarations}}\\
\noindent{\large Competing interests}\\
The authors declare no competing interests.

\noindent{\bf{\large Supporting Information}}\\
Additional experimental and theoretical details, including the procedure for sample growth, additional XPS-PEEM and LEED measurements on Co- and Se-terminated surfaces, EDS and XPS with large spot size to characterize the overall sample condition, ARPES constant energy maps on the two terminated surfaces to highlight the behavior of the surface states in the $k$-distribution, discussion on the spatial uniformity of terminations in relation to the analysis of micro-ARPES data, and the description of DFT modeling of the slabs, including the choice of supercell parameters.

\noindent{\bf{\large \centerline{References}}}

\bibliography{alter.bib}

@article{Edwards2023,
	Author = {Edwards, B. and Dowinton, O. and Hall, A. E. and Murgatroyd, P. A. E. and Buchberger, S. and Antonelli, T. and Siemann, G. -R. and Rajan, A. and Morales, E. Abarca and Zivanovic, A. and Bigi, C. and Belosludov, R. V. and Polley, C. M. and Carbone, D. and Mayoh, D. A. and Balakrishnan, G. and Bahramy, M. S. and King, P. D. C.},
	Da = {2023/04/01},
	Doi = {10.1038/s41563-022-01459-z},
	Id = {Edwards2023},
	Isbn = {1476-4660},
	Journal = {Nature Materials},
	Number = {4},
	Pages = {459--465},
	Title = {Giant valley-{Z}eeman coupling in the surface layer of an intercalated transition metal dichalcogenide},
	Ty = {JOUR},
	Url = {https://doi.org/10.1038/s41563-022-01459-z},
	Volume = {22},
	Year = {2023}}

@article{Mazin2022,
	Author = {Mazin, Igor},
	Collaboration = {The PRX Editors},
	Doi = {10.1103/PhysRevX.12.040002},
	Issue = {4},
	Journal = {Phys. Rev. X},
	Month = {Dec},
	Numpages = {3},
	Pages = {040002},
	Publisher = {American Physical Society},
	Title = {Editorial: Altermagnetism -- A New Punch Line of Fundamental Magnetism},
	Url = {https://link.aps.org/doi/10.1103/PhysRevX.12.040002},
	Volume = {12},
	Year = {2022}}

@article{Regmi2024,
	Title = {Altermagnetism in the layered intercalated transition metal dichalcogenide CoNb$_4$Se$_8$},
	Author = {Resham Babu Regmi and Hari Bhandari and Bishal Thapa and Yiqing Hao and Nileema Sharma and James McKenzie and Xinglong Chen and Abhijeet Nayak and Mohamed El Gazzah and Bence Gábor Márkus and László Forró and Xiaolong Liu and Huibo Cao and J. F. Mitchell and I. I. Mazin and Nirmal J. Ghimire},
	Journal = {Nat. Commun.},
    volume = {16},
    pages = {4399},
    year = {2024},
	Doi = {10.1038/s41467-025-58642-4},
	Url = { https://doi.org/10.1038/s41467-025-58642-4},
}

@article{Smejkal2022_1,
	Author = {\ifmmode \check{S}\else \v{S}\fi{}mejkal, Libor and Sinova, Jairo and Jungwirth, Tomas},
	Doi = {10.1103/PhysRevX.12.031042},
	Issue = {3},
	Journal = {Phys. Rev. X},
	Month = {Sep},
	Numpages = {16},
	Pages = {031042},
	Publisher = {American Physical Society},
	Title = {Beyond Conventional Ferromagnetism and Antiferromagnetism: A Phase with Nonrelativistic Spin and Crystal Rotation Symmetry},
	Url = {https://link.aps.org/doi/10.1103/PhysRevX.12.031042},
	Volume = {12},
	Year = {2022}}

@article{Smejkal2022_2,
	Author = {\ifmmode \check{S}\else \v{S}\fi{}mejkal, Libor and Sinova, Jairo and Jungwirth, Tomas},
	Doi = {10.1103/PhysRevX.12.040501},
	Issue = {4},
	Journal = {Phys. Rev. X},
	Month = {Dec},
	Numpages = {27},
	Pages = {040501},
	Publisher = {American Physical Society},
	Title = {Emerging Research Landscape of Altermagnetism},
	Url = {https://link.aps.org/doi/10.1103/PhysRevX.12.040501},
	Volume = {12},
	Year = {2022}}

@article{Bai2024,
    author = {Bai, Ling and Feng, Wanxiang and Liu, Siyuan and Šmejkal, Libor and Mokrousov, Yuriy and Yao, Yugui},
    title = {Altermagnetism: Exploring New Frontiers in Magnetism and Spintronics},
    journal = {Advanced Functional Materials},
    volume = {34},
    number = {49},
    pages = {2409327},
    doi = {https://doi.org/10.1002/adfm.202409327},
    url = {https://advanced.onlinelibrary.wiley.com/doi/abs/10.1002/adfm.202409327},
    year = {2024},
}

@article{Feng2022,
	Author = {Feng, Zexin and Zhou, Xiaorong and {\v S}mejkal, Libor and Wu, Lei and Zhu, Zengwei and Guo, Huixin and Gonz{\'a}lez-Hern{\'a}ndez, Rafael and Wang, Xiaoning and Yan, Han and Qin, Peixin and Zhang, Xin and Wu, Haojiang and Chen, Hongyu and Meng, Ziang and Liu, Li and Xia, Zhengcai and Sinova, Jairo and Jungwirth, Tom{\'a}{\v s} and Liu, Zhiqi},
	Da = {2022/11/01},
	Doi = {10.1038/s41928-022-00866-z},
	Id = {Feng2022},
	Isbn = {2520-1131},
	Journal = {Nature Electronics},
	Number = {11},
	Pages = {735--743},
	Title = {An anomalous {H}all effect in altermagnetic ruthenium dioxide},
	Ty = {JOUR},
	Url = {https://doi.org/10.1038/s41928-022-00866-z},
	Volume = {5},
	Year = {2022}}

@article{Fedchenko2024,
	author = {Olena Fedchenko  and Jan Minár  and Akashdeep Akashdeep  and Sunil Wilfred D’Souza  and Dmitry Vasilyev  and Olena Tkach  and Lukas Odenbreit  and Quynh Nguyen  and Dmytro Kutnyakhov  and Nils Wind  and Lukas Wenthaus  and Markus Scholz  and Kai Rossnagel  and Moritz Hoesch  and Martin Aeschlimann  and Benjamin Stadtmüller  and Mathias Kläui  and Gerd Schönhense  and Tomas Jungwirth  and Anna Birk Hellenes  and Gerhard Jakob  and Libor Šmejkal  and Jairo Sinova  and Hans-Joachim Elmers },
    title = {Observation of time-reversal symmetry breaking in the band structure of altermagnetic {RuO\textsubscript{2}}},
    journal = {Science Advances},
    volume = {10},
    number = {5},
    pages = {eadj4883},
    year = {2024},
    doi = {10.1126/sciadv.adj4883},
    URL = {https://www.science.org/doi/abs/10.1126/sciadv.adj4883},
}

@article{Krempasky2024,
	Author = {Krempask{\'y}, J. and {\v S}mejkal, L. and D'Souza, S. W. and Hajlaoui, M. and Springholz, G. and Uhl{\'\i}{\v r}ov{\'a}, K. and Alarab, F. and Constantinou, P. C. and Strocov, V. and Usanov, D. and Pudelko, W. R. and Gonz{\'a}lez-Hern{\'a}ndez, R. and Birk Hellenes, A. and Jansa, Z. and Reichlov{\'a}, H. and {\v S}ob{\'a}{\v n}, Z. and Gonzalez Betancourt, R. D. and Wadley, P. and Sinova, J. and Kriegner, D. and Min{\'a}r, J. and Dil, J. H. and Jungwirth, T.},
	Da = {2024/02/01},
	Doi = {10.1038/s41586-023-06907-7},
	Id = {Krempask{\'y}2024},
	Isbn = {1476-4687},
	Journal = {Nature},
	Number = {7999},
	Pages = {517--522},
	Title = {Altermagnetic lifting of {K}ramers spin degeneracy},
	Ty = {JOUR},
	Url = {https://doi.org/10.1038/s41586-023-06907-7},
	Volume = {626},
	Year = {2024}}

@article{Mazin2023,
	Author = {Mazin, I. I.},
	Doi = {10.1103/PhysRevB.107.L100418},
	Issue = {10},
	Journal = {Phys. Rev. B},
	Month = {Mar},
	Numpages = {4},
	Pages = {L100418},
	Publisher = {American Physical Society},
	Title = {Altermagnetism in {MnTe}: Origin, predicted manifestations, and routes to detwinning},
	Url = {https://link.aps.org/doi/10.1103/PhysRevB.107.L100418},
	Volume = {107},
	Year = {2023}}

@article{Turek2022,
	Author = {Turek, Ilja},
	Doi = {10.1103/PhysRevB.106.094432},
	Issue = {9},
	Journal = {Phys. Rev. B},
	Month = {Sep},
	Numpages = {14},
	Pages = {094432},
	Publisher = {American Physical Society},
	Title = {Altermagnetism and magnetic groups with pseudoscalar electron spin},
	Url = {https://link.aps.org/doi/10.1103/PhysRevB.106.094432},
	Volume = {106},
	Year = {2022}}

@article{Sattigeri2023,
	Author = {Sattigeri, Raghottam M. and Cuono, Giuseppe and Autieri, Carmine},
	Doi = {10.1039/D3NR03681B},
	Issue = {42},
	Journal = {Nanoscale},
	Pages = {16998-17005},
	Publisher = {The Royal Society of Chemistry},
	Title = {Altermagnetic surface states: towards the observation and utilization of altermagnetism in thin films{,} interfaces and topological materials},
	Url = {http://dx.doi.org/10.1039/D3NR03681B},
	Volume = {15},
	Year = {2023}}

@misc{Lin2024,
	Archiveprefix = {arXiv},
	Author = {Zihan Lin and Dong Chen and Wenlong Lu and Xin Liang and Shiyu Feng and Kohei Yamagami and Jacek Osiecki and Mats Leandersson and Balasubramanian Thiagarajan and Junwei Liu and Claudia Felser and Junzhang Ma},
	Eprint = {2402.04995},
	Primaryclass = {cond-mat.mtrl-sci},
	Title = {Observation of Giant Spin Splitting and d-wave Spin Texture in Room Temperature Altermagnet {RuO2}},
	Url = {https://arxiv.org/abs/2402.04995},
	Year = {2024}}

@article{QE-2017,
  author={P Giannozzi and O Andreussi and T Brumme and O Bunau and M Buongiorno Nardelli
  and M Calandra and R Car and C Cavazzoni and D Ceresoli and M Cococcioni and N Colonna
  and I Carnimeo and A Dal Corso and S de Gironcoli and P Delugas and R A DiStasio Jr and A Ferretti
  and A Floris and G Fratesi and G Fugallo and R Gebauer and U Gerstmann and F Giustino and T Gorni
  and J Jia and M Kawamura and H-Y Ko and A Kokalj and E Küçükbenli and M Lazzeri and M Marsili
  and N Marzari and F Mauri and N L Nguyen and H-V Nguyen and A Otero-de-la-Roza and L Paulatto
  and S Poncé and D Rocca and R Sabatini and B Santra and M Schlipf and A P Seitsonen
  and A Smogunov and I Timrov and T Thonhauser and P Umari and N Vast and X Wu and S Baroni},
  title={Advanced capabilities for materials modelling with {QUANTUM ESPRESSO}},
  journal={Journal of Physics: Condensed Matter},
  volume={29},
  number={46},
  pages={465901},
  url={http://stacks.iop.org/0953-8984/29/i=46/a=465901},
  year={2017},
}

@article{PhysRevB.85.085201,
  title = {Extracting \textit{E} versus $\vec{k}$ effective band structure from supercell calculations on alloys and impurities},
  author = {Popescu, Voicu and Zunger, Alex},
  journal = {Phys. Rev. B},
  volume = {85},
  issue = {8},
  pages = {085201},
  numpages = {12},
  year = {2012},
  month = {Feb},
  publisher = {American Physical Society},
  doi = {10.1103/PhysRevB.85.085201},
  url = {https://link.aps.org/doi/10.1103/PhysRevB.85.085201},
}

@article{Yuan2023,
author = {Yuan, Lin-Ding and Zunger, Alex},
title = {Degeneracy Removal of Spin Bands in Collinear Antiferromagnets with Non-Interconvertible Spin-Structure Motif Pair},
journal = {Advanced Materials},
volume = {35},
number = {31},
pages = {2211966},
doi = {https://doi.org/10.1002/adma.202211966},
url = {https://advanced.onlinelibrary.wiley.com/doi/abs/10.1002/adma.202211966},
eprint = {https://advanced.onlinelibrary.wiley.com/doi/pdf/10.1002/adma.202211966},
year = {2023}
}

@article{DeVita2026b,
  title = {Uncovering surface states of the Dirac semimetal ${\mathrm{BaMg}}_{2}{\mathrm{Bi}}_{2}$},
  author = {De Vita, A. and Bakkelund, J. and \ifmmode \acute{S}\else \'{S}\fi{}wiatek, H. and Winiarski, M. J. and Malick, S. and Nielsen, C. V. B. and Bertran, F. and Jones, A. J. H. and Majchrzak, P. and Granozio, F. Miletto and Miwa, J. A. and Ernstorfer, R. and Pincelli, T. and Klimczuk, T. and Bigi, C. and Mazzola, F.},
  journal = {Phys. Rev. B},
  volume = {113},
  issue = {3},
  pages = {035148},
  numpages = {6},
  year = {2026},
  month = {Jan},
  publisher = {American Physical Society},
  doi = {10.1103/ps93-rfk3},
  url = {https://link.aps.org/doi/10.1103/ps93-rfk3}
}

@article{Liu2026,
  title = {Symmetry-driven spin splitting in altermagnets: an angle-resolved photoemission spectroscopy perspective},
  author = {Liu, J and Ma, X. and Zhang, X. and Jing, W. and Liu, Z. and Shen, D.},
  journal = {Nano Convergence},
  volume = {13},
  issue = {6},
  year = {2026},
  doi = {10.1186/s40580-026-00536-2},
  url = {https://doi.org/10.1186/s40580-026-00536-2}
}

@article{Liu2024,
  title = {Absence of Altermagnetic Spin Splitting Character in Rutile Oxide ${\mathrm{RuO}}_{2}$},
  author = {Liu, Jiayu and Zhan, Jie and Li, Tongrui and Liu, Jishan and Cheng, Shufan and Shi, Yuming and Deng, Liwei and Zhang, Meng and Li, Chihao and Ding, Jianyang and Jiang, Qi and Ye, Mao and Liu, Zhengtai and Jiang, Zhicheng and Wang, Siyu and Li, Qian and Xie, Yanwu and Wang, Yilin and Qiao, Shan and Wen, Jinsheng and Sun, Yan and Shen, Dawei},
  journal = {Phys. Rev. Lett.},
  volume = {133},
  issue = {17},
  pages = {176401},
  numpages = {7},
  year = {2024},
  month = {Oct},
  publisher = {American Physical Society},
  doi = {10.1103/PhysRevLett.133.176401},
  url = {https://link.aps.org/doi/10.1103/PhysRevLett.133.176401}
}

@misc{Devita2026,
      title={Robust spin splitting and fermiology in a layered altermagnet}, 
      author={Alessandro De Vita and Chiara Bigi and Davide Romanin and Matthew D. Watson and Vincent Polewczyk and Marta Zonno and François Bertran and My Bang Petersen and Federico Motti and Giovanni Vinai and Manuel Tuniz and Federico Cilento and Mario Cuoco and Brian M. Andersen and Andreas Kreisel and Luciano Jacopo D'Onofrio and Oliver J. Clark and Mark T. Edmonds and Christopher Candelora and Muxian Xu and Siyu Cheng and Alexander LaFleur and Tommaso Antonelli and Giorgio Sangiovanni and Lorenzo Del Re and Ivana Vobornik and Jun Fujii and Fabio Miletto Granozio and Alessia Sambri and Emiliano Di Gennaro and Jeppe B. Jacobsen and Henrik Jacobsen and Iulia Cojocariu and Marcin Szpytma and Andrea Locatelli and Tevfik Mentes and Matthieu Jamet and Jean-François Jacquot and Pasquale Orgiani and Ralph Ernstorfer and Ilija Zeljkovic and Younghun Hwang and Matteo Calandra and Jill A. Miwa and Federico Mazzola},
      year={2026},
      eprint={2502.20010},
      archivePrefix={arXiv},
      primaryClass={cond-mat.str-el},
      url={https://arxiv.org/abs/2502.20010}, 
}

@article{Candelora_2026,
	author = {Candelora, Christopher and Xu, Muxian and Cheng, Siyu and De Vita, Alessandro and Romanin, Davide and Bigi, Chiara and Petersen, My Bang and LaFleur, Alexander and Castro, Wilber Alfaro and Motti, Federico and Orgiani, Pasquale and Vinai, Giovanni and Calandra, Matteo and Miwa, Jill and Hwang, Younghun and Wang, Ziqiang and Mazzola, Federico and Zeljkovic, Ilija},
	date = {2026/01/31},
	doi = {10.1038/s43246-026-01081-5},
	id = {Candelora2026},
	isbn = {2662-4443},
	journal = {Communications Materials},
	title = {Discovery of magnetic-field-tunable density modulations and spin tilting in a layered altermagnet},
	url = {https://doi.org/10.1038/s43246-026-01081-5},
	year = {2026}}

@article{Mazzola_2023_co,
	author = {Mazzola, Federico and Enzner, Stefan and Eck, Philipp and Bigi, Chiara and Jugovac, Matteo and Cojocariu, Iulia and Feyer, Vitaliy and Shu, Zhixue and Pierantozzi, Gian Marco and De Vita, Alessandro and Carrara, Pietro and Fujii, Jun and King, Phil D. C. and Vinai, Giovanni and Orgiani, Pasquale and Cacho, Cephise and Watson, Matthew D. and Rossi, Giorgio and Vobornik, Ivana and Kong, Tai and Di Sante, Domenico and Sangiovanni, Giorgio and Panaccione, Giancarlo},
	date = {2023/09/13},
	doi = {10.1021/acs.nanolett.3c02022},
	isbn = {1530-6984},
	journal = {Nano Letters},
	journal1 = {Nano Letters},
	journal2 = {Nano Lett.},
	month = {09},
	number = {17},
	pages = {8035--8042},
	publisher = {American Chemical Society},
	title = {Observation of Termination-Dependent Topological Connectivity in a Magnetic Weyl Kagome Lattice},
	type = {doi: 10.1021/acs.nanolett.3c02022},
	url = {https://doi.org/10.1021/acs.nanolett.3c02022},
	volume = {23},
	year = {2023},
	year1 = {2023}}

@article{DiSante_2023,
	author = {Di Sante, Domenico and Bigi, Chiara and Eck, Philipp and Enzner, Stefan and Consiglio, Armando and Pokharel, Ganesh and Carrara, Pietro and Orgiani, Pasquale and Polewczyk, Vincent and Fujii, Jun and King, Phil D. C. and Vobornik, Ivana and Rossi, Giorgio and Zeljkovic, Ilija and Wilson, Stephen D. and Thomale, Ronny and Sangiovanni, Giorgio and Panaccione, Giancarlo and Mazzola, Federico},
	date = {2023/08/01},
	doi = {10.1038/s41567-023-02053-z},
	id = {Di Sante2023},
	isbn = {1745-2481},
	journal = {Nature Physics},
	number = {8},
	pages = {1135--1142},
	title = {Flat band separation and robust spin Berry curvature in bilayer kagome metals},
	url = {https://doi.org/10.1038/s41567-023-02053-z},
	volume = {19},
	year = {2023}}

@article{
Mazzola_2018,
author = {Federico Mazzola  and Veronika Sunko  and Seunghyun Khim  and Helge Rosner  and Pallavi Kushwaha  and Oliver J. Clark  and Lewis Bawden  and Igor Markovic  and Timur K. Kim  and Moritz Hoesch  and Andrew P. Mackenzie  and Phil D. C. King },
title = {Itinerant ferromagnetism of the Pd-terminated polar surface of PdCoO<sub>2</sub>},
journal = {Proceedings of the National Academy of Sciences},
volume = {115},
number = {51},
pages = {12956-12960},
year = {2018},
doi = {10.1073/pnas.1811873115},
URL = {https://www.pnas.org/doi/abs/10.1073/pnas.1811873115},
eprint = {https://www.pnas.org/doi/pdf/10.1073/pnas.1811873115}}

@article{Mazzola_2024,
	author = {Mazzola, Federico and Brzezicki, Wojciech and Mercaldo, Maria Teresa and Guarino, Anita and Bigi, Chiara and Miwa, Jill A. and De Fazio, Domenico and Crepaldi, Alberto and Fujii, Jun and Rossi, Giorgio and Orgiani, Pasquale and Chaluvadi, Sandeep Kumar and Chalil, Shyni Punathum and Panaccione, Giancarlo and Jana, Anupam and Polewczyk, Vincent and Vobornik, Ivana and Kim, Changyoung and Miletto-Granozio, Fabio and Fittipaldi, Rosalba and Ortix, Carmine and Cuoco, Mario and Vecchione, Antonio},
	date = {2024/02/01},
	doi = {10.1038/s41586-024-07033-8},
	id = {Mazzola2024},
	isbn = {1476-4687},
	journal = {Nature},
	number = {8000},
	pages = {752--758},
	title = {Signatures of a surface spin--orbital chiral metal},
	url = {https://doi.org/10.1038/s41586-024-07033-8},
	volume = {626},
	year = {2024}}

@article{Bigi_2021,
  title = {Measuring spin-polarized electronic states of quantum materials: $2H\ensuremath{-}{\mathrm{NbSe}}_{2}$},
  author = {Bigi, Chiara and Mazzola, Federico and Fujii, Jun and Vobornik, Ivana and Panaccione, Giancarlo and Rossi, Giorgio},
  journal = {Phys. Rev. B},
  volume = {103},
  issue = {24},
  pages = {245142},
  numpages = {9},
  year = {2021},
  month = {Jun},
  publisher = {American Physical Society},
  doi = {10.1103/PhysRevB.103.245142},
  url = {https://link.aps.org/doi/10.1103/PhysRevB.103.245142}
}

@article{Bawden_2016,
	author = {Bawden, L. and Cooil, S. P. and Mazzola, F. and Riley, J. M. and Collins-McIntyre, L. J. and Sunko, V. and Hunvik, K. W. B. and Leandersson, M. and Polley, C. M. and Balasubramanian, T. and Kim, T. K. and Hoesch, M. and Wells, J. W. and Balakrishnan, G. and Bahramy, M. S. and King, P. D. C.},
	date = {2016/05/23},
	doi = {10.1038/ncomms11711},
	id = {Bawden2016},
	isbn = {2041-1723},
	journal = {Nature Communications},
	number = {1},
	pages = {11711},
	title = {Spin--valley locking in the normal state of a transition-metal dichalcogenide superconductor},
	url = {https://doi.org/10.1038/ncomms11711},
	volume = {7},
	year = {2016}}

@article{QEcode,

  author={Paolo Giannozzi and Stefano Baroni and Nicola Bonini and Matteo Calandra and Roberto Car and Carlo Cavazzoni and Davide

Ceresoli and Guido L Chiarotti and Matteo Cococcioni and Ismaila Dabo and Andrea Dal Corso and Stefano de

Gironcoli and Stefano Fabris and Guido Fratesi and Ralph Gebauer and Uwe Gerstmann and Christos Gougoussis and Anton

Kokalj and Michele Lazzeri and Layla Martin-Samos and Nicola Marzari and Francesco Mauri and Riccardo Mazzarello and Stefano

Paolini and Alfredo Pasquarello and Lorenzo Paulatto and Carlo Sbraccia and Sandro Scandolo and Gabriele Sclauzero and Ari P

Seitsonen and Alexander Smogunov and Paolo Umari and Renata M Wentzcovitch},

  title={QUANTUM ESPRESSO: a modular and open-source software project for quantum simulations of materials},

  journal={Journal of Physics: Condensed Matter},

  volume={21},

  number={39},

  pages={395502},

  url={http://stacks.iop.org/0953-8984/21/i=39/a=395502},

  year={2009}

}

@article{QE-2020,

  doi = {10.1063/5.0005082},

  url = {https://doi.org/10.1063/5.0005082},

  year = {2020},

  month = apr,

  publisher = {{AIP} Publishing},

  volume = {152},

  number = {15},

  pages = {154105},

  author = {Paolo Giannozzi and Oscar Baseggio and Pietro Bonf{\`{a}} and Davide Brunato and Roberto Car and Ivan Carnimeo and Carlo Cavazzoni and Stefano de Gironcoli and Pietro Delugas and Fabrizio Ferrari Ruffino and Andrea Ferretti and Nicola Marzari and Iurii Timrov and Andrea Urru and Stefano Baroni},

  title = {Quantum ESPRESSO toward the exascale},

  journal = {The Journal of Chemical Physics}

}

@article{PBE,

  title = {Generalized Gradient Approximation Made Simple},

  volume = {77},

  ISSN = {1079-7114},

  url = {http://dx.doi.org/10.1103/PhysRevLett.77.3865},

  DOI = {10.1103/physrevlett.77.3865},

  number = {18},

  journal = {Physical Review Letters},

  publisher = {American Physical Society (APS)},

  author = {Perdew,  John P. and Burke,  Kieron and Ernzerhof,  Matthias},

  year = {1996},

  month = oct,

  pages = {3865–3868}

}

@article{Methfessel1989,

  title = {High-precision sampling for Brillouin-zone integration in metals},

  volume = {40},

  ISSN = {0163-1829},

  url = {http://dx.doi.org/10.1103/PhysRevB.40.3616},

  DOI = {10.1103/physrevb.40.3616},

  number = {6},

  journal = {Physical Review B},

  publisher = {American Physical Society (APS)},

  author = {Methfessel,  M. and Paxton,  A. T.},

  year = {1989},

  month = aug,

  pages = {3616–3621}

}

@article{Pacil2021,

  title = {Narrowing of d bands of FeCo layers intercalated under graphene},

  volume = {118},

  ISSN = {1077-3118},

  url = {http://dx.doi.org/10.1063/5.0047266},

  DOI = {10.1063/5.0047266},

  number = {12},

  journal = {Applied Physics Letters},

  publisher = {AIP Publishing},

  author = {Pacilè,  Daniela and Cardoso,  Claudia and Avvisati,  Giulia and Vobornik,  Ivana and Mariani,  Carlo and Leon,  Dario A. and Bonfà,  Pietro and Varsano,  Daniele and Ferretti,  Andrea and Betti,  Maria Grazia},

  year = {2021},

  month = mar

}

\clearpage
\setcounter{figure}{0}
\setcounter{equation}{0}
\setcounter{table}{0}
\renewcommand{\thefigure}{S\arabic{figure}}
\renewcommand{\theequation}{S\arabic{equation}}
\renewcommand{\thetable}{S\arabic{table}}
\renewcommand{\theHfigure}{S\arabic{figure}}
\renewcommand{\theHequation}{S\arabic{equation}}
\renewcommand{\theHtable}{S\arabic{table}}

\begin{center}
{\large\bfseries Supporting Information for\\[2pt]
``Termination-dependent surface Fermiology in the intercalated altermagnet CoNb$_4$Se$_8$''}
\end{center}
\vspace{1em}

\subsection{Growth details}
High-purity Co (99.99$\%$), Nb (99.999$\%$), and Se (99.9999$\%$) powders were used as starting materials. Prior to loading, the quartz ampoule was chemically cleaned and vacuum-annealed to minimise contamination and residual oxygen. The elements were sealed in an evacuated quartz ampoule (10\,mm diameter, 150\,mm length) together with iodine (5\,mg/cm$^3$ relative to ampoule volume) serving as the transport agent. Crystal growth was carried out in a two-zone horizontal furnace, with the source zone kept at a higher temperature than the growth zone. For CoNb$_4$Se$_8$, the source region was maintained at 960–980\,$^\circ$C, while the growth region temperature was gradually increased from 880\,$^\circ$C to 900\,$^\circ$C over 100 hours. Both zones were then held at constant temperature for an additional 300 hours to promote the formation of large single crystals. The furnace was subsequently cooled over 100 hours to 200\,$^\circ$C (source) and 100\,$^\circ$C (growth), after which the ampoule was allowed to return to room temperature naturally. Typical crystals obtained under these conditions measured approximately 5\,$\times$\,5\,$\times$\,0.1\,mm$^3$. Residual iodine was removed by rinsing the crystals in methanol. The Co concentration $x$ in Co$_x$NbSe$_2$ was first estimated using energy-dispersive x-ray spectroscopy (EDS) in a field-emission scanning electron microscope. This methodology is the same shown in Ref.\cite{Devita2026}. Samples' compositions were monitored by using XPS, as shown in the main text, and EDS from scanning electron microscope. Their magnetic properties where also determined by SQUID as shown in our previous work in Ref.\cite{Devita2026}.\\

\subsection{Core level analysis and terminations}

We analyse the change in the XPS intensity ratio between Se and Co core levels measured by micro-ARPES and reported in Fig.~\ref{fig:1} of the main text. The core level spectra were fitted using Lorentzian line shapes, together with a linear background, in order to extract the integrated peak intensities. This yields
\begin{equation}
R_0=\left(\frac{I_{\mathrm{Se}}}{I_{\mathrm{Co}}}\right)_0=2.41589,
\qquad
R=\left(\frac{I_{\mathrm{Se}}}{I_{\mathrm{Co}}}\right)=1.79734.
\end{equation}

Assuming that the screening of the Se signal due to the overlying Co in the Co-terminated surface is negligible, the variation of the ratio reflects a change in the Co photoemission intensity. The relative increase of the Co signal is therefore
\begin{equation}
\frac{I_{\mathrm{Co}}}{I_{\mathrm{Co},0}}=\frac{R_0}{R}
=\frac{2.41589}{1.79734}
=1.34415,
\end{equation}
which corresponds to
\begin{equation}
\frac{\Delta I_{\mathrm{Co}}}{I_{\mathrm{Co},0}}=0.34415,
\end{equation}
i.e.\ a $34.4\%$ enhancement of the Co signal.

To relate this increase to a surface Co coverage, we employ a standard exponential attenuation model. The photoemission intensity from a periodic stack of equivalent Co-containing planes at depths $nd$ is
\begin{equation}
I_{\mathrm{Co},0} \propto \sum_{n=1}^{\infty} e^{-nd/\Lambda}
=
\frac{e^{-d/\Lambda}}{1-e^{-d/\Lambda}},
\end{equation}
where $d$ is the spacing between successive Co planes along the surface normal, and
\begin{equation}
\Lambda=\lambda\cos\theta
\end{equation}
is the effective attenuation length, with $\lambda$ the inelastic mean free path and $\theta$ the emission angle.

We then introduce an additional Co contribution located in the outermost surface layer. The surface coverage is denoted by $\Theta$, expressed in monolayers, where one fully occupied allowed Co surface layer corresponds to $0.25$ ML. The corresponding intensity contribution is
\begin{equation}
I_{\mathrm{surf}} \propto \frac{\Theta}{0.25},
\end{equation}
since it is not attenuated by overlying material.

The total Co signal becomes
\begin{equation}
I_{\mathrm{Co}} = I_{\mathrm{Co},0} + I_{\mathrm{surf}},
\end{equation}
and the ratio between the modified and reference intensities is
\begin{equation}
\frac{I_{\mathrm{Co}}}{I_{\mathrm{Co},0}}
=
1+\frac{I_{\mathrm{surf}}}{I_{\mathrm{Co},0}}
=
1+\frac{\Theta/0.25}{e^{-d/\Lambda}/(1-e^{-d/\Lambda})}.
\end{equation}

This can be rewritten as
\begin{equation}
\frac{I_{\mathrm{Co}}}{I_{\mathrm{Co},0}}
=
1+\frac{\Theta}{0.25}\left(e^{d/\Lambda}-1\right).
\end{equation}

Using the experimental value $I_{\mathrm{Co}}/I_{\mathrm{Co},0}=1.34415$, we obtain
\begin{equation}
1+\frac{\Theta}{0.25}\left(e^{d/\Lambda}-1\right)=1.34415,
\end{equation}
which yields
\begin{equation}
\Theta
=
0.25\,\frac{0.34415}{e^{d/\Lambda}-1}.
\end{equation}

For typical XPS conditions, the first Co plane lies at a small fraction of an attenuation length below the surface, such that
\begin{equation}
\frac{d}{\Lambda}\approx 0.28\text{--}0.33.
\end{equation}

Inserting this range gives
\begin{equation}
\Theta \approx 0.22\text{--}0.27\ \mathrm{ML},
\end{equation}
with a central value
\begin{equation}
\Theta \approx 0.25\ \mathrm{ML}.
\end{equation}

\begin{figure}
\includegraphics[width=\columnwidth]{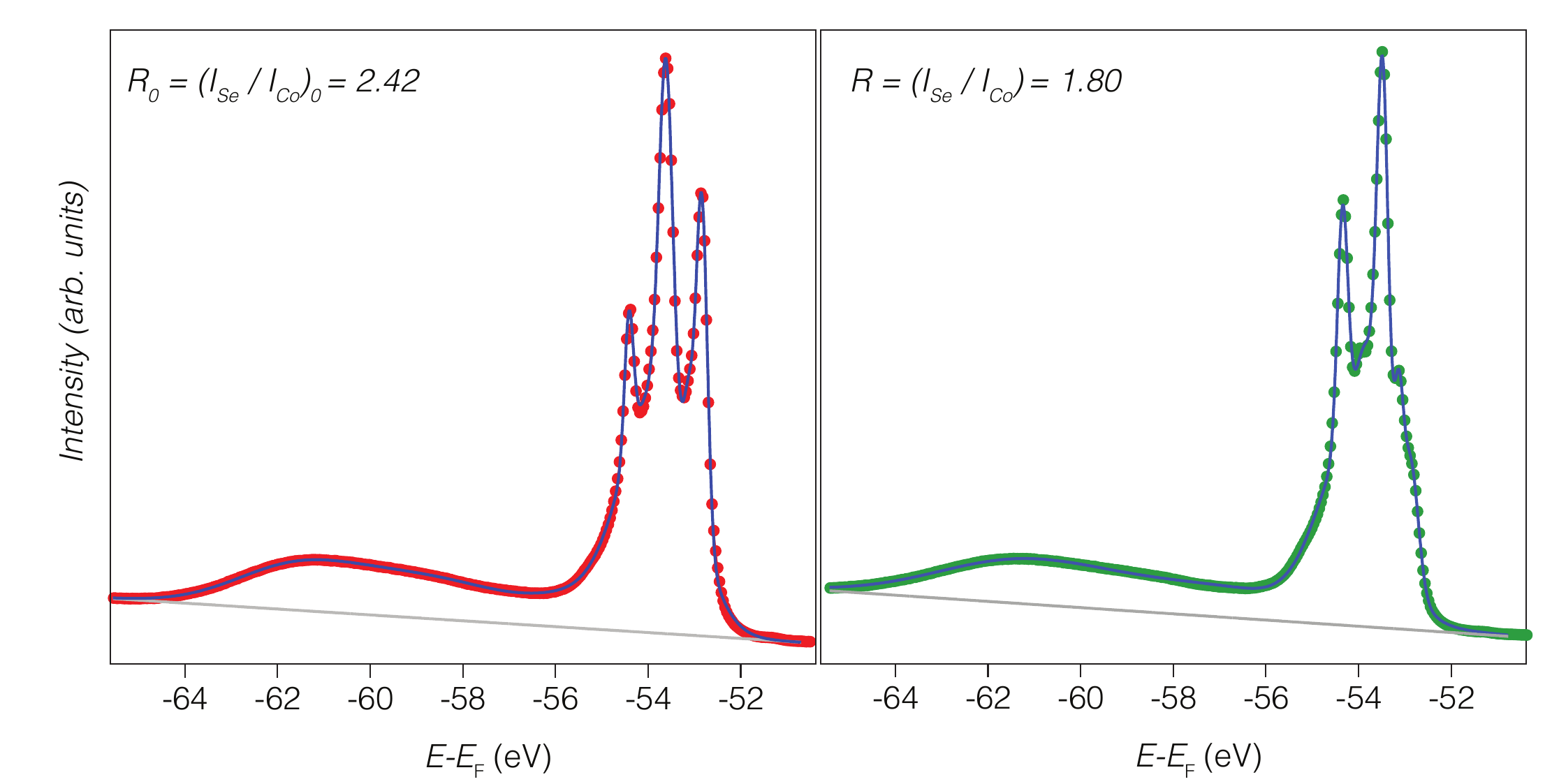}
\caption{XPS core-level spectra used to extract the Se/Co intensity ratios. Experimental data (symbols) are fitted with Lorentzian line shapes, including a linear background (grey line), with the total fit shown as solid lines. Left: reference spectrum yielding $R_0 = (I_{\mathrm{Se}}/I_{\mathrm{Co}})_0 = 2.42$. Right: modified surface spectrum yielding $R = (I_{\mathrm{Se}}/I_{\mathrm{Co}}) = 1.80$. The reduction of the Se/Co ratio reflects an enhanced Co contribution at the surface.}
  \label{fig:S1}
\end{figure}

The decrease of the XPS Se/Co intensity ratio from $2.41589$ to $1.79734$ thus corresponds to a $34.4\%$ enhancement of the Co signal and is quantitatively consistent with an almost fully occupied allowed Co surface layer. The result of this analysis is shown in Fig.\ref{fig:S1}. Note that this procedure can only be considered with a large uncertainty, specifically associated to the mean free path variation and also to the possibility of additional Se peaks tailing off into the Co region, being these very close to each-other. Thus, despite the tentative quantification, the aim of this analysis is to rather give a qualitative trend.

\begin{figure}
\includegraphics[width=\columnwidth]{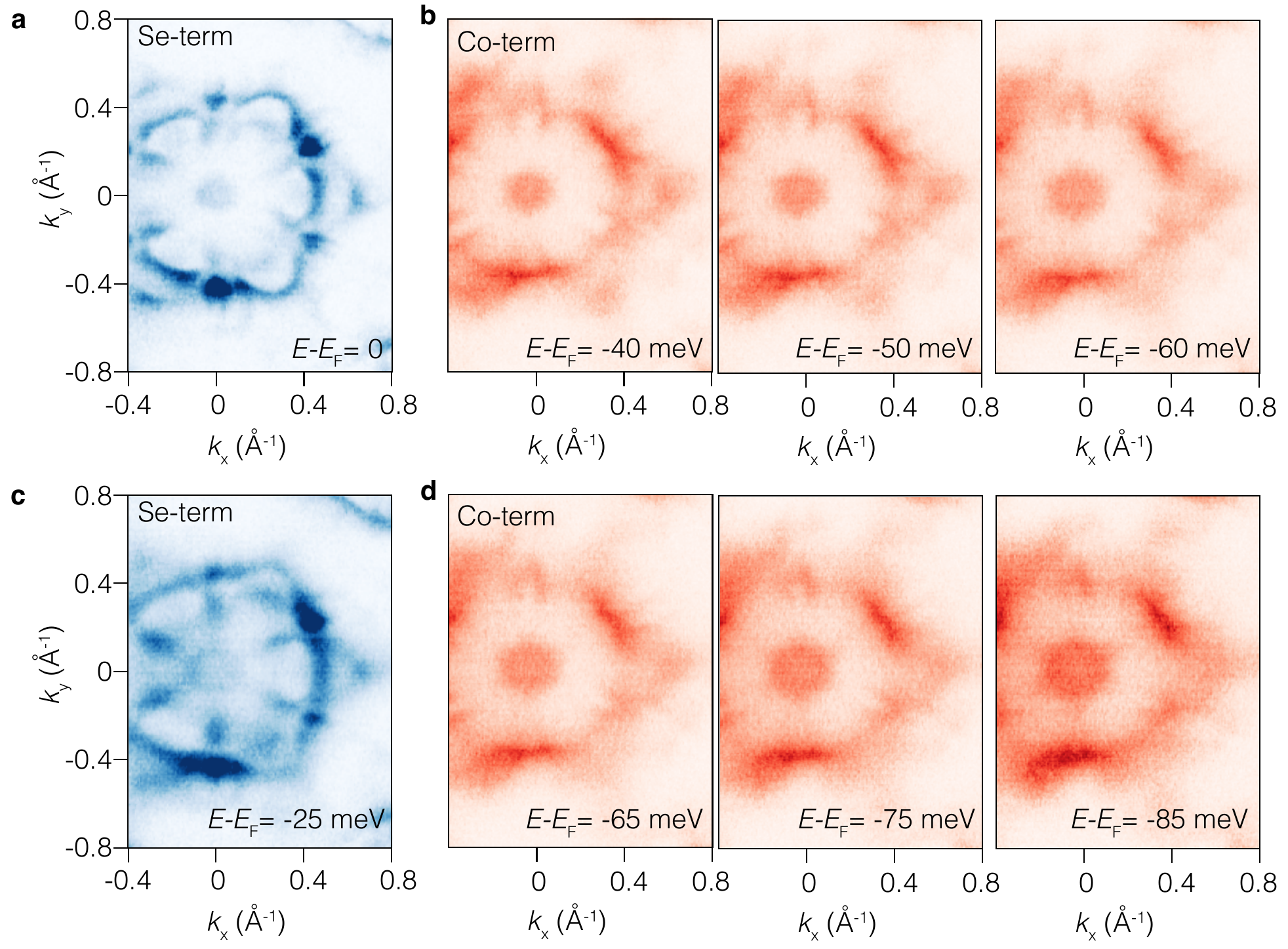}
  \caption{Constant-energy maps acquired by micro-ARPES for the two surface terminations. \textbf{a.} Se-terminated surface at $E-E_F=\SI{0}{\eV}$. \textbf{b.} Co-terminated surface at $E-E_F=\SI{-40}{\milli\eV}$, \SI{-50}{\milli\eV}, and \SI{-60}{\milli\eV}. \textbf{c.} Se-terminated surface at $E-E_F=\SI{-25}{\milli\eV}$. \textbf{d.} Co-terminated surface at $E-E_F=\SI{-65}{\milli\eV}$, $\SI{-75}{\milli\eV}$, and $\SI{-85}{\milli\eV}$. The first binding-energy offset between the two terminations is chosen to approximately compensate the $\sim\SI{40}{\milli\eV}$ chemical-potential shift discussed in the main text, and the other panels want to provide an additional comparison. While the central pocket area can be matched under this shift, the remaining features do not overlap, demonstrating that the distinct Fermiologies of the two terminations cannot be explained by a rigid energy displacement.}
  \label{fig:S2}
\end{figure}

\subsection{Constant energy maps and binding energy comparison}

Figure~\ref{fig:S2} reports a series of constant-energy maps for the Se- and Co-terminated surfaces, acquired under identical experimental conditions. The purpose of this analysis is to test whether the differences observed in the Fermi surfaces of the two terminations can be reconciled by a simple rigid chemical-potential shift. For the Se-terminated surface (Fig.~\ref{fig:S2}a), the Fermi surface at $E-E_F=\SI{0}{\milli\eV}$ exhibits a flower-like contour with pronounced petal-shaped segments. Upon moving to $E-E_F=\SI{-25}{\milli\eV}$, the central pocket increases in size, consistently with a hole-like dispersion. For the Co-terminated surface (Fig.~\ref{fig:S2}b), analogous constant-energy maps are shown at $E-E_F=\SI{-40}{\milli\eV}$, $\SI{-50}{\milli\eV}$, and \SI{-60}{\milli\eV}. As discussed in the main text, the Co termination exhibits an overall shift of approximately \SI{40}{\milli\eV} relative to the Se termination. For completeness, Fig.~\ref{fig:S2}c compares the Se-terminated map at $E-E_F=\SI{-25}{\milli\eV}$ with Co-terminated maps at $E-E_F=\SI{-65}{\milli\eV}$, \SI{-75}{\milli\eV}, and \SI{-85}{\milli\eV}, corresponding to an additional $\sim\SI{-60}{\milli\eV}$ downward shift.

Under a purely rigid energy displacement, matching the area of the central pocket should simultaneously align the remaining dispersive features, preserving their number and connectivity. While the central pocket size can indeed be approximately matched, the outer petal-like structures and associated intensity segments do not superimpose between the two terminations. Their shape, symmetry distribution, and relative spectral weight remain qualitatively distinct even after compensating for the chemical-potential offset. This analysis unambiguously demonstrates that the termination-dependent Fermi surfaces cannot be obtained from one another by a uniform shift in $E_F$. Instead, the two terminations host intrinsically different surface electronic structures, consistent with the slab DFT calculations presented in Fig.~\ref{fig:3} of the main text.

\subsection{Broadening effects in real experiment}

In angle-resolved photoemission, the out-of-plane momentum $k_z$ is not a good quantum number of the photoelectron final state due to the finite escape depth $\lambda$. As a consequence, the experiment does not probe a single $k_z$, but rather a distribution centred at $k_{z0}$ with a finite width
\begin{equation}
\Delta k_z \sim \frac{1}{\lambda}.
\end{equation}

The measured photocurrent at fixed in-plane momentum $k_\parallel$ is therefore effectively a weighted average over a range of $k_z$,
\begin{equation}
I(k_\parallel,\omega) \propto \int dk_z\, W(k_z-k_{z0})\, A(k_\parallel,k_z,\omega),
\end{equation}
where $A(k_\parallel,k_z,\omega)$ is the spectral function and $W$ is a broad function of width $\Delta k_z$.

This effect becomes particularly pronounced in materials with a large lattice parameter $c$, as in the case of CoNb$_4$Se$_8$ for which the Brillouin zone height is reduced to

\begin{equation}
\frac{2\pi}{c}.
\end{equation}

In this case, even a moderate $\Delta k_z$ can represent a substantial fraction of the full $k_z$ Brillouin zone, i.e.
\begin{equation}
\frac{\Delta k_z}{2\pi/c} \not\ll 1,
\end{equation}
so that the measured signal contains contributions from a wide range of inequivalent $k_z$ values.

For bands with finite out-of-plane dispersion, $E = E(k_\parallel,k_z)$, this averaging leads to an intrinsic spectral broadening of order
\begin{equation}
\Delta E \sim \left|\frac{\partial E}{\partial k_z}\right| \Delta k_z,
\end{equation}
and, more generally, to a redistribution of spectral weight reflecting the underlying $k_z$-dependent dispersion. In the limiting case where $\Delta k_z$ spans a significant fraction of the Brillouin zone, the observed spectrum approaches a $k_z$-integrated signal rather than a well-defined $k_z$ cut, as reflected in our experimental data.

\subsection{Other possible sources of broadening}

The finite linewidth observed in the ARPES spectra can arise from several intrinsic and extrinsic mechanisms. From a many-body perspective, electron-electron interactions lead to a finite quasiparticle lifetime, resulting in an energy-dependent broadening of the spectral function. In addition, electron-phonon coupling provides an additional scattering channel, which can further renormalise the dispersion and contribute to the linewidth, particularly at finite temperature. Scattering from residual impurities or disorder acts as an approximately energy-independent contribution, leading to a nearly constant background broadening of the bands.

Beyond these intrinsic effects, extrinsic factors related to the surface quality are expected to play a significant role. In particular, for a Co-terminated surface of CoNb$_4$Se$_8$, the cleaving process may preferentially remove or displace Co atoms, which are comparatively less strongly bound. This can result in a finite density of defects and vacancies at the surface, giving rise to additional scattering channels. Such disorder-induced scattering leads to a predominantly homogeneous broadening of the spectral features. Furthermore, deviations from the ideal surface stoichiometry, such as Co deficiency, may locally modify the electronic potential and charge balance, thereby slightly shifting the chemical potential. As a consequence, the exact position of the Fermi level with respect to the calculated electronic structure may also be affected.

\subsection{DFT details}

Density Functional Theory (DFT) calculations were carried out within the collinear spin–polarized framework using the plane–wave pseudopotential approach as implemented in the \textsc{Quantum ESPRESSO} package~\cite{QEcode,QE-2017,QE-2020}. The electron–ion interaction was described using a norm–conserving pseudopotential for Se atoms, while ultrasoft pseudopotentials were adopted for Nb and Co atoms. The Perdew–Burke–Ernzerhof (PBE) generalized gradient approximation was used for the exchange–correlation functional~\cite{PBE}. A kinetic–energy cutoff of 45~Ry for the wavefunctions and 540~Ry for the charge density was employed.

Integration over the Brillouin zone of the bulk Co–doped $2\times2$ NbSe$_2$ supercell in the altermagnetic phase was performed using a $9\times9\times16$ Monkhorst–Pack $k$–point grid together with a first–order Methfessel–Paxton smearing of 0.005~Ry~\cite{Methfessel1989}. The resulting electronic dispersion was subsequently unfolded into the primitive $1\times1$ NbSe$_2$ Brillouin zone through the unfolding procedure implemented in the \texttt{unfold.x} code~\cite{PhysRevB.85.085201,Pacil2021}. The spectral weight, defined as the projection of the supercell eigenstates onto the Bloch states of the primitive cell at fixed $\mathbf{k}$, is used to construct the unfolded electronic dispersion. The continuous unfolded spectral-intensity maps and calculated Fermi-surface images were generated on an energy grid with step $\Delta E=0.005~\mathrm{eV}$ by replacing each eigenvalue $E_n$ with a Lorentzian spectral function $L(E-E_n)=\pi^{-1}w[(E-E_n)^2+w^2]^{-1}$, using a broadening parameter $w=0.005~\mathrm{eV}$. The unfolding was performed from the $2\times2\times1$ supercell to the primitive Brillouin zone. To evaluate surface effects with different terminations we constructed slab geometries. For the Se termination we employed a slab consisting of five NbSe$_2$ layers with four Co atoms and a vacuum region of \SI{45}{\angstrom} along the non–periodic direction perpendicular to the surface. For the Co termination we used a slab composed of three NbSe$_2$ layers with four Co atoms and a vacuum region of \SI{55}{\angstrom}. The slab thickness was chosen in order to preserve the altermagnetic configuration while simultaneously allowing the realization of both surface terminations. It is worth noting that, because of the unfolding into the primitive zone, which is hexagonal, the calculated electronic structure also displays a hexagonal symmetry pattern. At the surface, however, a real experiment would capture the effect of symmetry breaking, resulting in a three-fold symmetry, as observed in the constant-energy maps measured by micro-ARPES.

For the slab calculations the atomic positions were fully relaxed by increasing the kinetic–energy cutoff to 60~Ry for the wavefunctions and 720~Ry for the charge density. Structural optimization was performed until the residual forces on each atom were smaller than $10^{-4}$~Ry/Bohr. The ground–state electronic density was obtained using a $9\times9\times1$ $k$–point mesh with a first–order Methfessel–Paxton smearing of 0.005~Ry. Atom-projected DOS curves were obtained from the relaxed slab calculations by projecting the Kohn--Sham states onto Nb, Se, and Co atomic contributions for each termination. To assess the surface-versus-bulk character of the low-energy spectral weight, we also compared projected DOS from the outer surface layers with DOS from inner, bulk-like layers of the slab.

\subsection{DFT bulk versus surface states}

\begin{figure}
\includegraphics[width=0.5\columnwidth]{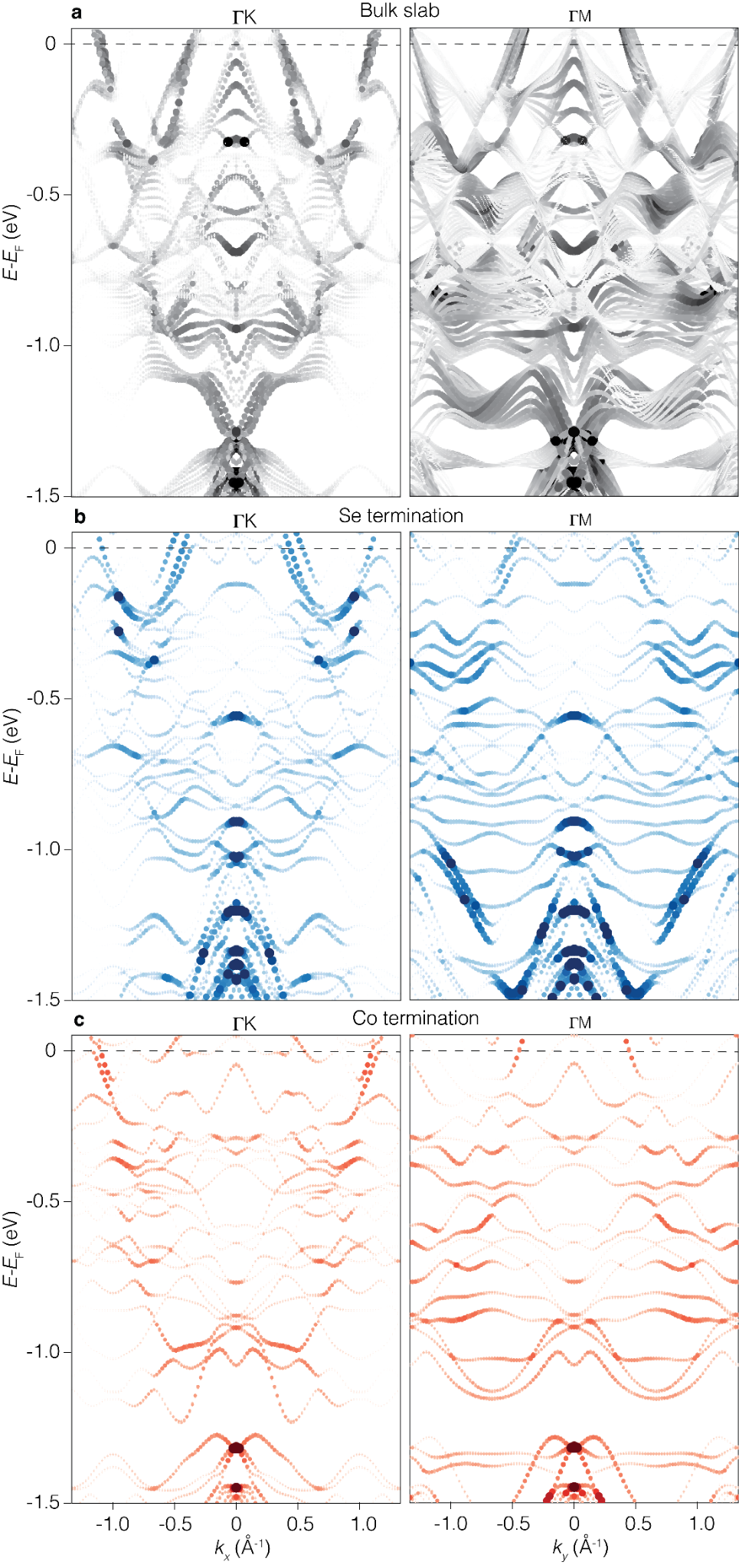}
  \caption{DFT electronic structure comparison between bulk and slab geometries.  (a) Bulk band structure projected along the $\bar{\Gamma}-\bar{K}$ and $\bar{\Gamma}-\bar{M}$ directions, showing the continuum of bulk states.  (b) Slab calculation for the Se-terminated surface, where additional weakly dispersive states appear below $E_F$, particularly along $\bar{\Gamma}-\bar{M}$.  (c) Slab calculation for the Co-terminated surface, exhibiting a distinct set of termination-dependent bands shifted to higher binding energy and with modified dispersion. The comparison distinguishes features spectrally separated from the projected bulk continuum, which are classified as surface-localised states, from termination-dependent features overlapping the continuum, which are described as surface resonances.}
  \label{fig:S3}
\end{figure}

To disentangle the origin of the electronic states observed experimentally, we compare bulk and slab DFT calculations, explicitly accounting for the two possible surface terminations.

The bulk electronic structure is characterised by dispersive bands with finite $k_z$ dependence, reflecting the three-dimensional periodicity of the crystal. When projected onto the surface Brillouin zone, these states form a continuum of spectral weight. This bulk continuum is used as the reference against which the slab bands are classified.

The slab calculations reveal termination-dependent additional spectral weight. We classify a slab feature as an isolated surface-localised state only when it appears in an energy--momentum region where no bulk band is present in the bulk projection and when its dispersion or spectral weight depends on the termination. If a termination-dependent feature overlaps the projected bulk continuum, we instead refer to it as a surface resonance. This distinction avoids assigning a strictly surface-localised character to features, such as the central $\bar{\Gamma}$-centred pocket, whose spectral weight can overlap the bulk projection.

For the Se-terminated slab, the calculations show weakly dispersive bands located just below $E_F$, particularly evident along the $\bar{\Gamma}-\bar{M}$ direction. These states form relatively flat features that are termination dependent but partly embedded within the projected bulk continuum, and are therefore best described as surface resonances in the overlapping regions. For the Co-terminated slab, the corresponding states are shifted to higher binding energy and acquire a distinct diamond-like dispersion, indicating a different surface potential and hybridisation.

Along the $\bar{\Gamma}-\bar{K}$ direction, both terminations exhibit surface-derived bands that approach and cross the Fermi level, forming the characteristic petal- and spindle-like features observed in the calculated Fermi surfaces. Where these features are spectrally separated from the projected bulk continuum, the slab calculations support their identification as surface-localised states.

A direct comparison between bulk and slab calculations shows that the termination-dependent features cannot be obtained from the bulk band structure by a rigid energy shift or simple projection. Instead, they arise from the modified surface potential and from surface--bulk hybridisation, with their detailed shape and energy position controlled by the termination.

These results demonstrate that the low-energy electronic structure near the Fermi level contains a significant contribution from termination-dependent surface-derived states and surface resonances. The central $\bar{\Gamma}$-centred pocket is used in the main text only as a registration feature for the rigid-shift analysis and is not used as evidence for the isolated surface-state assignment of the other bands.

\subsection{Spatial uniformity of spectroscopic signals}

To further expand from Fig.~\ref{fig:1} of the main text the discussion on the spatial variation of surface terminations, we have plotted micro-ARPES spectra acquired at several different positions of the spatial map, as shown in Figure~\ref{fig:fig1NewRep} below.

The spatial maps in Fig.~\ref{fig:1}b of the main text and in Fig.~\ref{fig:fig1NewRep} are constructed by raster-scanning the micro-ARPES beam across the sample. At each spatial coordinate $(x,y)$, the local fingerprint intensity is obtained by integrating the measured photoemission intensity over a fixed energy--momentum window,
\begin{equation}
I_{\mathrm{fp}}(x,y)=\iint_W I(x,y;k,E)\, dk\, dE, \qquad W=\{E_1\le E\le E_2,\; k_1\le k\le k_2\}.
\end{equation}
The window is selected to isolate a termination-specific valence-band fingerprint identified from the slab-DFT calculations and from representative single-domain spectra, rather than a generic chemical contrast. For the map shown here, an example valence-band window is $E\in[43.0,43.13]~\mathrm{eV}$ and $k\in[-1.85,-0.85]~\mathrm{\AA}^{-1}$. A complementary core-level window in the Se $3d$/Co $3p$ region was used as a chemical check, but not as the sole criterion for assignment, because core-level maps can include bulk contributions and can retain intensity from tilted or poorly ordered regions. We did not apply an automated intensity-ratio threshold to assign the spots. Instead, the map was used as a practical spatial guide to locate candidate domains, and each selected position was accepted only when the manually focused micro-ARPES spectrum showed both the appropriate core-level lineshape and the sharp near-$E_F$ termination fingerprint. Tilted or disordered regions were rejected because they lacked the well-centred, coherent dispersion of a single termination.

\begin{figure}
    \centering
    \includegraphics[width=0.75\linewidth]{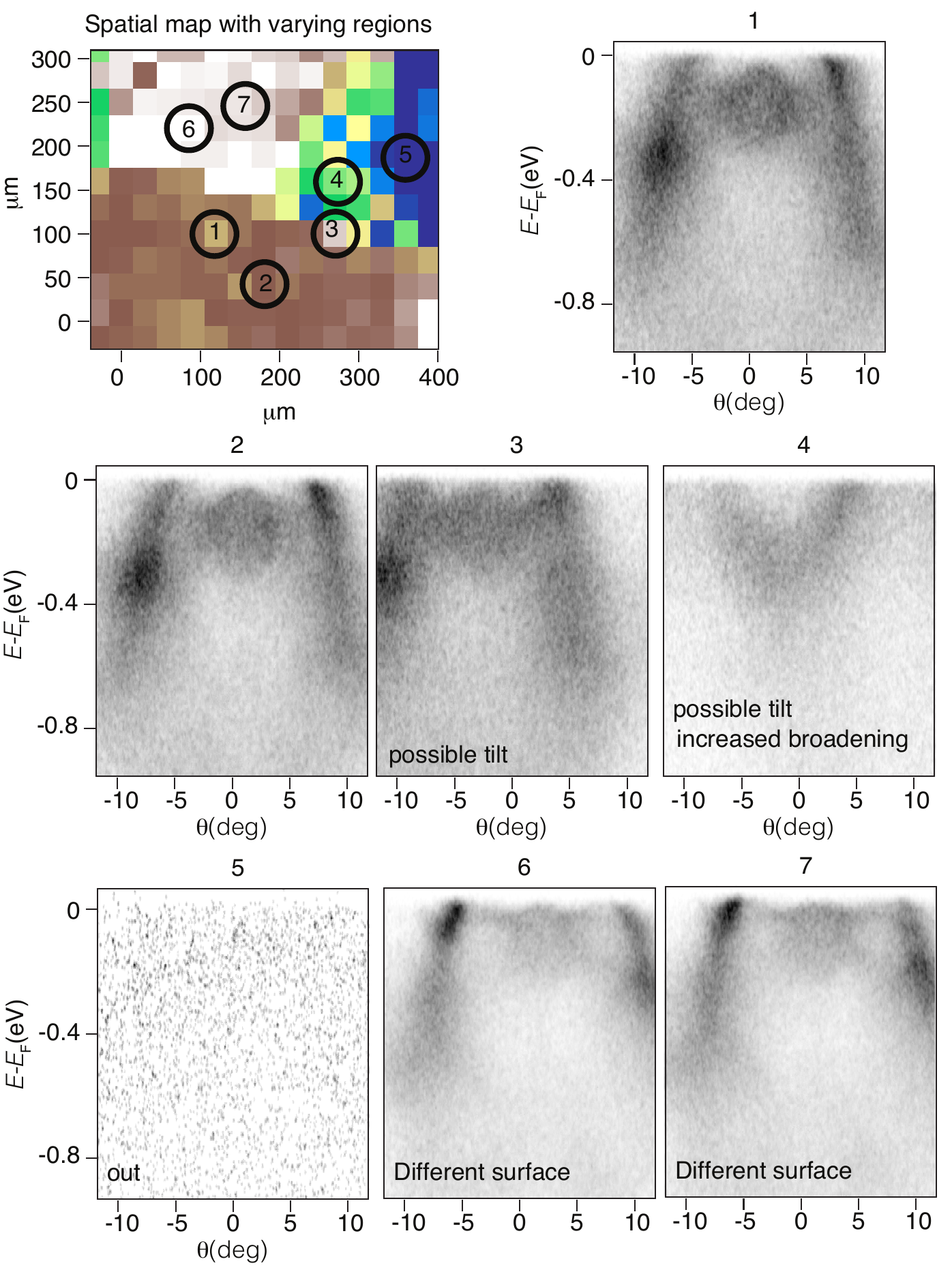}
    \caption{Spatially-resolved ARPES investigation. The top-left panel is a spatial map ($\approx$ 400$\times$\SI{300}{\micro\metre}) obtained by integrating the valence-band intensity over a termination-sensitive window, for example $E\in[43.0,43.13]~\mathrm{eV}$ and $k\in[-1.85,-0.85]~\mathrm{\AA}^{-1}$. The colour therefore encodes a spectral fingerprint used to guide the search for single-termination regions, not an automated intensity-threshold classification. Seven positions are circled (1-7), and the other panels show the micro-ARPES (\textit{E}, \textit{k}) dispersions taken at each of those points, plotting $E-E_F$ vs emission angle $\theta$ over roughly $\pm\ang{10}$.}
    \label{fig:fig1NewRep}
\end{figure}

Positions 1 and 2, in the homogeneous brown–yellow region, show clean, coherent dispersive bands — the well-ordered single-termination spectra. Each displayed spectrum is a single micro-ARPES acquisition from the circled position and is not itself an integrated spatial map. For the high-resolution measurements the beamline conditions were optimized spot-by-spot, including analyser focus, beam focus, slits, and acquisition conditions, and the data were acquired in swept mode to suppress detector artefacts. No additional algorithmic treatment was used for the termination assignment. Position 3 looks very similar but with a slight distortion we attribute to a possible tilt of the flake and a slightly larger disorder, as corroborated by an increased broadening of the spectra. Position 4 shows the same onset of tilt together with noticeably increased broadening, consistent with approaching a less well-ordered area. Position 5, in the dark-blue corner at the edge, is essentially featureless noise — labelled "out," \textit{i.e.}, the beam is out of the sample. Positions 6 and 7, in the bright-white region, show a clearly different dispersion ("different surface") — the other termination.

Figure~\ref{fig:fig1NewRep} makes the order/disorder-plus-termination picture explicit and visual: a central well-ordered single-termination domain (1, 2) giving sharp coherent bands; a gradient toward the edges where flake tilting and the onset of disorder progressively broaden and distort the spectra (3, 4); a point off the sample with no signal (5); and a spatially distinct second termination (6, 7) with its own dispersion. This directly supports that coherent spectra come from ordered single-termination patches, whereas by approaching the edges the signal degrades through tilt and disorder rather than through a smooth sub-stoichiometric Co gradient.

\newpage
\subsection{Atom- and layer-projected density of states}

We added atom-projected DOS calculations for both slab terminations to clarify the microscopic character of the near-$E_F$ spectral weight. The states close to $E_F$ have mixed atomic character with termination-dependent Nb, Se, and Co contributions. In the Se-terminated slab, the Se-projected component remains sizeable near $E_F$, whereas in the Co-terminated slab the Co-projected component is enhanced near the Fermi level. We also compare the projected DOS of surface layers with inner, bulk-like layers of the slab. This comparison shows that the near-$E_F$ spectral weight contains both surface-enhanced and bulk-like contributions, supporting the distinction used in the main text between isolated surface-localised states and termination-dependent surface resonances.

\begin{figure}
    \centering
    \includegraphics[width=0.86\linewidth]{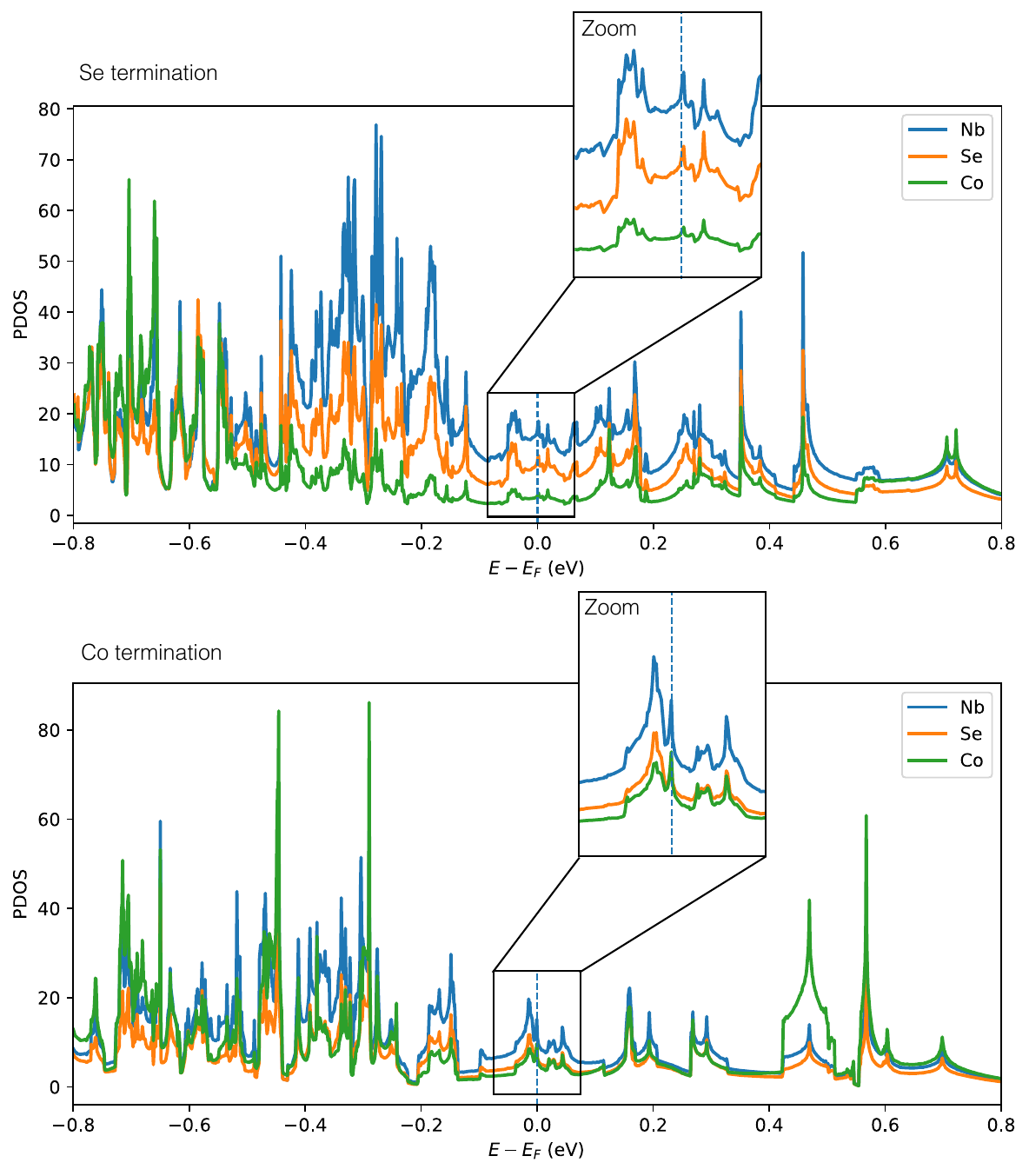}
    \caption{Atom-projected density of states for the Se-terminated and Co-terminated slab geometries. The PDOS is decomposed into Nb, Se, and Co contributions, with zoomed insets around $E_F$. The states close to $E_F$ have mixed atomic character, with sizeable Se-projected weight for the Se termination and enhanced Co-projected weight for the Co termination.}
    \label{fig:pdos_terminations}
\end{figure}

\begin{figure}
    \centering
    \includegraphics[width=0.86\linewidth]{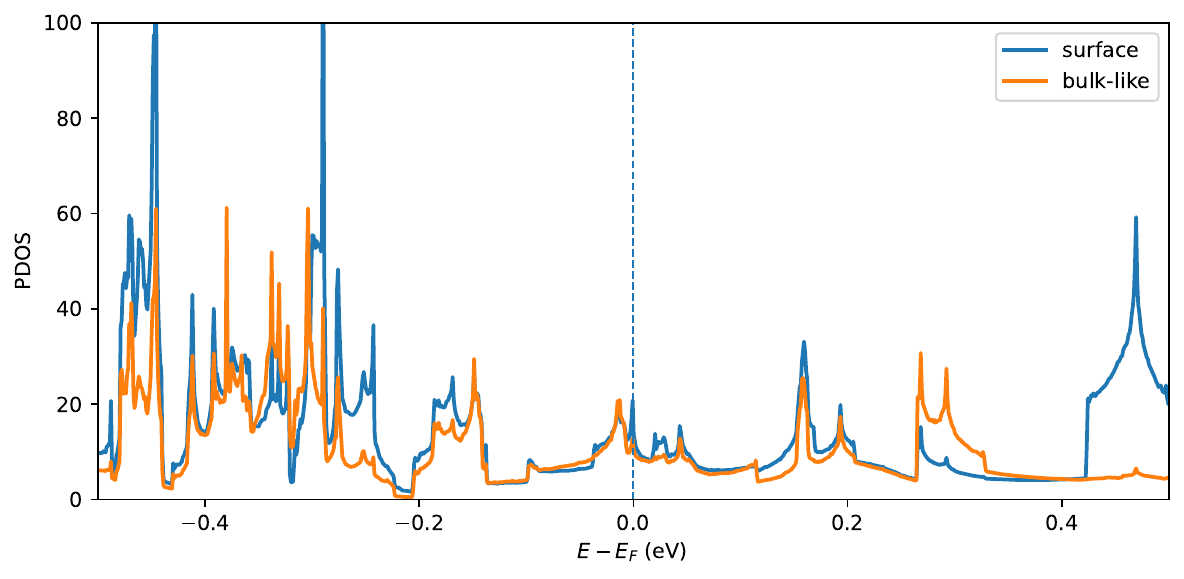}
    \caption{Surface-layer and bulk-like projected density of states from the slab calculation. The comparison shows that the low-energy spectral weight includes both enhanced surface-layer contributions and bulk-like components near $E_F$, consistent with describing overlapping termination-dependent features as surface resonances rather than strictly isolated surface states.}
    \label{fig:surface_vs_bulk_pdos}
\end{figure}

\end{document}